\documentclass[a4paper,aps,preprintnumbers,showpacs,twocolumn,superscriptaddress,nofootinbib,amsmath,amssymb]{revtex4-1}
\usepackage[dvips]{graphics}
\usepackage{color,hyperref}
\usepackage{times}
\hypersetup{
  colorlinks=true,        % false: boxed links; true: colored links
  linkcolor=blue,         % color of internal links
  citecolor=magenta,      % color of links to bibliography
  filecolor=magenta,      % color of file links
  urlcolor=red            % color of external links
}

\newcommand{\cf}{{cf.,}~}
\newcommand{\ie}{{i.e.,}~}
\newcommand{\eg}{{e.g.,}~}

\def\acond{\left\vert_{\begin{array}{l}\scriptstyle r=r_0\\\scriptstyle\epsilon_1=\epsilon_2=0\end{array}}\right.\hskip-0.1cm}
\def\bcond{\left\vert_{\begin{array}{l}\scriptstyle r=r_0\\\scriptstyle\epsilon_1=0\end{array}}\right.\hskip-0.1cm}

\begin{document}
\title{New parametrization for spherically symmetric black holes in
  metric theories of gravity}

\author{Luciano Rezzolla}
\affiliation{Institut f\"ur Theoretische Physik, Goethe-Universit\"at,
  Max-von-Laue-Str. 1, 60438 Frankfurt, Germany}
\affiliation{Max-Planck-Institut f\"ur Gravitationsphysik, Albert
  Einstein Institut, Am M\"uhlenberg 1, 14476 Potsdam, Germany}
\author{Alexander Zhidenko}
\affiliation{Institut f\"ur Theoretische Physik, Goethe-Universit\"at,
  Max-von-Laue-Str. 1, 60438 Frankfurt, Germany}
\affiliation{Centro de Matem\'atica, Computa\c{c}\~ao e Cogni\c{c}\~ao,
  Universidade Federal do ABC (UFABC), Rua Aboli\c{c}\~ao, CEP:
  09210-180, Santo Andr\'e, SP, Brazil}

\begin{abstract}
We propose a new parametric framework to describe in generic metric
theories of gravity the spacetime of spherically symmetric and slowly
rotating black holes. In contrast to similar approaches proposed so far, we do not
use a Taylor expansion in powers of $M/r$, where $M$ and $r$ are the mass
of the black hole and a generic radial coordinate, respectively. Rather,
we use a continued-fraction expansion in terms of a compactified radial
coordinate. This choice leads to superior convergence properties and
allows us to approximate a number of known metric theories with a much
smaller set of coefficients. The measure of these coefficients via
observations of near-horizon processes can be used to effectively
constrain and compare arbitrary metric theories of gravity. Although our
attention is here focussed on spherically symmetric black holes, we also
discuss how our approach could be extended to rotating black holes.
\end{abstract}
\pacs{04.50.Kd,04.70.Bw,04.25.Nx,04.30.-w,04.80.Cc}
\maketitle

\section{Introduction}

Black holes are one of the most intriguing, fascinating and yet
unsettling consequences of classical general relativity. Even when
putting aside the acceptance and understanding of the physical
singularities hidden at their centers, the mere existence of an event
horizon leads to a number of unsolved problems and long-standing
debates. Yet, black holes are some of the most cherished objects in
modern astronomy and evidence of their existence at different scales
appears as common as it is convincing.

Proof of the existence of an event horizon would not be disputable if it
appeared in terms of gravitational radiation, for instance in the form of
a quasinormal mode ringdown when a new black hole is formed. However, it
would become surely difficult, if possible at all, when using the
electromagnetic emission coming from material accreting onto
it~\cite{Abramowicz:02}. At the same time, our increasing ability to
perform astronomical observations that probe regions on scales that are
comparable or even smaller than the size of the event horizon, will soon
put us in the position of posing precise questions on the physical
properties of those astronomical objects that appear to have all the
properties of black holes in general relativity.

A good example in this respect is offered by astronomical observations of
the radio compact source Sgr~A$^*$, which resides at the center of our
Galaxy and is commonly assumed to be a supermassive black hole. Recent
radio observations of Sgr~A$^*$ have been made on scales comparable to
what would be the size of the event horizon if it indeed were a black
hole~\cite{Doeleman}. Furthermore, in the near future, very long baseline
interferometric radio observations are expected to image the so-called
black-hole ``shadow''~\cite{Falcke:00}, namely the photon ring marking
the surface where photons on circular orbits will have their smallest
stable orbit \cite{Johannsen:2012vz}. These observations, besides
providing the long-sought evidence for the existence of black holes, will
also provide the possibility of testing the no-hair theorem in general
relativity \cite{Bambi:2008jg,JohannsenPsaltis:2010,Loeb:2013lfa}.

If sufficiently accurate, the planned astronomical observations will not
only provide convincing evidence for the existence of an event horizon,
but they will also indicate if deviations exist from the predictions of
general relativity. However, given the already large number of
alternative theories of gravity, and considering that this is only
expected to grow in the near future, a case-by-case validation of a given
theory using the observational data does not seem as viable an option. It
is instead much more reasonable to develop a model-independent framework
that parametrizes the most generic black-hole geometry through a finite
number of adjustable quantities. These quantities must be chosen in such
a way that they can be used to measure deviations from general relativity
(or a black-hole geometry) and, at the same time, can be estimated
robustly from the observational data \cite{Vigeland:2011ji}.

This approach is not particularly new and actually similar in spirit to
the parametrized post-Newtonian approach (PPN) developed in the 1970s to
describe the dynamics of binary systems of compact
stars~\cite{Will:2005va}. A first step in this direction was done by
Johannsen and Psaltis~\cite{Johannsen:2011dh}, who have proposed a
general expression for the metric of a spinning non-Kerr black hole in
which the deviations from general relativity are expressed in terms of a
Taylor expansion in powers of $M/r$, where $M$ and $r$ are the mass of
the black hole and a generic radial coordinate. While some of the first
coefficients of the expansion can be easily constrained in terms of
PPN-like parameters, an infinite number remains to be determined from
observations near the event horizon \cite{Johannsen:2011dh}. This
approach was recently generalized by relaxing the area-mass relation for
non-Kerr black holes and introducing two independent modifications of the
metric functions $g_{tt}$ and $g_{rr}$
\cite{Cardoso:2014rha}. Unfortunately, as discussed in
\cite{Cardoso:2014rha}, this approach can face some difficulties:

\begin{enumerate}
  \item The proposed metric is described by an infinite number of
    parameters, which are roughly equally important in the strong-field
    regime, making it difficult to isolate the dominant terms.

  \item The parametrization can be specialized to reproduce a spherically
    symmetric black hole metric in alternative theories only in the case
    in which the deviation from the general relativity is small. This was
    checked for the black holes in dilatonic Einstein-Gauss-Bonnet gravity
    \cite{Einstein-Dilaton-Gauss-Bonnet-black-hole}, for which the
    corresponding parameters were calculated only in the regime of small
    coupling.

  \item At first order in the spin, the parametrization cannot reproduce
    deviations from the Kerr metric arising in alternative theories of
    gravity. As an example, it cannot reproduce the modifications arising
    for a slowly rotating black hole in Chern-Simons modified gravity.

\end{enumerate}

In this paper we propose a solution to these issues and take another step
in the direction of deriving a general parametrization for objects in
metric theories of gravity. More precisely, we propose a parametrization
for spherically symmetric and slowly rotating black hole geometries which
can mimic black holes with a high accuracy and with a small number of
free coefficients. This is achieved by expressing the deviations from
general relativity in terms of a continued-fraction expansion via a
compactified radial coordinate defined between the event horizon and
spatial infinity. The superior convergence properties of this expansion
effectively reduces to a few the number of coefficients necessary to
approximate such spherically symmetric metric to the precision that can
be in principle probed with near-future observations.  While
phenomenologically effective, the approach we suggest has also an obvious
drawback. Because the metric expression we propose is not the consistent
result of any alternative theory of gravity, it does not have any
guarantee of being physically relevant or nothing more than a
mathematical exercise.

The paper is organized as follows. In Sec.~\ref{sec:param} we describe
the proposed parametrization method. Sec.~\ref{sec:JP} is devoted to the
relation between the proposed parameters and the parameters of the
Johannsen-Psaltis spherically symmetric black hole. In
Sec.~\ref{sec:dilaton} we obtain values of the parameters that
approximate a dilaton black hole, while in Sec.~\ref{sec:observe} we
compare the photon circular orbit, the innermost stable circular orbit,
and the quasinormal ringing predicted within our approximation with the
corresponding quantities obtained for the exact solution of a dilaton
black hole. In Sec.~\ref{sec:rotation} we apply our approach to slowly
rotating black holes and, in the conclusions, we discuss applications for
our framework and its possible generalization for the axisymmetric
case. Finally, Appendix~\ref{appendix_a} is dedicated to a comparison of
our parametrization framework with the alternative parametrization of a
spherically symmetric black hole proposed in Ref.~\cite{Cardoso:2014rha}.

\section{Parametrization of spherically symmetric black holes}
\label{sec:param}

The line element of any spherically symmetric stationary configuration in
a spherical polar coordinate system $(t,r,\theta,\phi)$ can be written as
\begin{equation}
\label{ssbh}
ds^2=-N^2(r)dt^2+\frac{B^2(r)}{N^2(r)}dr^2+r^2 d\Omega^2\,,
\end{equation}
where $d\Omega^2 \equiv d\theta^2+\sin^2\theta d\phi^2$, and $N$ and $B$
are functions of the radial coordinate $r$ only.

For any metric theory of gravity whose line element can be expressed as
\eqref{ssbh}, we will next require that it could contain a spherically
symmetric black hole\footnote{Much of what discussed here for a black
  hole can be employed also for the spacetime of a compact star. In this
  case, however, suitable boundary conditions for the metric will need to
  be imposed at the stellar surface $x=0$ [\cf
    Eq.~\eqref{eq:bh_hor}].}. By this we mean that the spacetime could
contain a surface where the expansion of radially outgoing photons is
zero, and define this surface as the event horizon. We mark its radial
position as $r = r_0 > 0$ and this definition implies that
\begin{equation}
\label{eq:bh_hor}
N(r_0)=0\,.
\end{equation}
Furthermore, we will neglect any cosmological effect, so that the
asymptotic properties of the line element \eqref{ssbh} will be those of
an asymptotically flat spacetime. Differently from previous approaches,
we find it convenient to compactify the radial coordinate and introduce
the dimensionless variable
\begin{equation}
x \equiv 1-\frac{r_0}{r}\,,
\end{equation}
so that $x=0$ corresponds to the location of the event horizon, while
$x=1$ corresponds to spatial infinity. In addition, we rewrite the metric
function $N$ as
\begin{equation}
\label{N2}
N^2=x A(x)\,,
\end{equation}
where
\begin{equation}
A(x)>0 \quad\mbox{for}\quad 0\leq x\leq1 \,.
\end{equation}
We further express the functions $A$ and $B$ after introducing three
additional terms, $\epsilon$, $a_0$, and $b_0$, so that
\begin{eqnarray}
\label{asympfix_1}
A(x)&=&1-\epsilon (1-x)+(a_0-\epsilon)(1-x)^2+{\tilde A}(x)(1-x)^3,\nonumber\\
\\
\label{asympfix_2}
B(x)&=&1+b_0(1-x)+{\tilde B}(x)(1-x)^2,
\end{eqnarray}
where the functions ${\tilde A}$ and ${\tilde B}$ are introduced to
describe the metric near the horizon (\ie for $x \simeq 0$) and are
finite there, as well as at spatial infinity (\ie for $x \simeq 1$).

Since we are not considering any specific theory of gravity, we do not
have precise constraints to impose on the metric functions $N$ and
$B$. At the same time, we can exploit the information deduced from the
PPN expansion to constrain their asymptotic expression, \ie their
behaviour for $x \simeq 1$~\cite{Will:2005va}. More specifically, we can
include the PPN asymptotic behaviour by expressing $B$ and $N$ as
\begin{align}
\label{expansion_1}
N^2 = 1&-\frac{2M}{r}+(\beta-\gamma)\frac{2M^2}{r^2}+
{\cal O}\left(r^{-3}\right)\nonumber\\
= 1& -\frac{2M}{r_0}(1-x)+(\beta-\gamma)\frac{2M^2}{r_0^2}(1-x)^2
\nonumber\\
&+{\cal O}\left((1-x)^3\right)\,,\\
\label{expansion_2}
\frac{B^2}{N^2} = 1& +\gamma\frac{2M}{r}+
{\cal O}\left(r^{-2}\right)=1+\gamma\frac{2M}{r_0}(1-x)\nonumber\\
&+{\cal O}\left((1-x)^2\right)\,.
\end{align}
Here $M$ is the Arnowitt-Deser-Misner (ADM) mass of the spacetime, while
$\beta$ and $\gamma$ are the PPN parameters, which are observationally
constrained to be~\cite{Will:2005va}
\begin{equation}\label{PPN}
|\beta-1|\lesssim2.3\times10^{-4},
\qquad|\gamma-1|\lesssim2.3\times10^{-5}
\,.
\end{equation}
Note that we have expanded the metric function $g_{tt}$ to ${\cal
  O}\left((1-x)^3\right)$, but $g_{rr}$ to ${\cal
  O}\left((1-x)^2\right)$. The reason for this difference is that the
highest-order PPN constraint on $g_{rr}$, \ie the parameter $\gamma$, is
at first order in $(1-x)$. Conversely, the parameters $\beta$ and
$\gamma$ set constraints on $g_{tt}$ at second order in $(1-x)$.

By comparing the two asymptotic expansions
\eqref{asympfix_1}--\eqref{asympfix_2} and
(\ref{expansion_1})--(\ref{expansion_2}), and collecting terms at the
same order, we find that
\begin{align}
 1+\epsilon &= \frac{2M}{r_0}\,, \\
 a_0 &= (\beta-\gamma)\frac{2M^2}{r_0^2}\,,\\
 1+\epsilon+2b_0 &=\gamma\frac{2M}{r_0}\,.
\end{align}
Hence, the introduced dimensionless constant $\epsilon$ is completely
fixed by the horizon radius $r_0$ and the ADM mass $M$ as
\begin{equation}
\epsilon=\frac{2M-r_0}{r_0} = - \left(1 - \frac{2M}{r_0}\right)\,,
\end{equation}
and thus measures the deviations of $r_0$ from $2M$. On the other hand,
the coefficients $a_0$ and $b_0$ can be seen as combinations of the PPN
parameters as
\begin{align}
&a_0=\frac{(\beta-\gamma)(1+\epsilon)^2}{2}\,, \\
&b_0=\frac{(\gamma-1)(1+\epsilon)}{2}\,.
\end{align}
or, alternatively, as
\begin{align}
&\beta  = 1 + \frac{2\left[a_0 + b_0(1+\epsilon)\right]}{(1+\epsilon)^{2}}\,,\\
&\gamma = 1 + \frac{2b_0}{1+\epsilon}\,.
\end{align}
Using now the observations constraints (\ref{PPN}) on the PPN parameters,
we conclude that $a_0$ and $b_0$ are both small and, in particular, $a_0
\sim b_0 \sim 10^{-4}$.

As mentioned above, the functions ${\tilde A}(x)$ and ${\tilde B}(x)$
have the delicate task of describing the black hole metric near its
horizon and should therefore have superior convergence properties than
those offered, for instance, by a simple Taylor expansion. We chose
therefore to express them in terms of rational functions (see also
Ref.~\cite{Aether}). Since the asymptotic behavior of the metric is fixed
by the conditions (\ref{asympfix_1})--(\ref{asympfix_2}), it is
convenient to parametrize ${\tilde A}(x)$ and ${\tilde B}(x)$ by Pad\'e
approximants in the form of continued fractions, \ie as
\begin{subequations}
\label{contfrac}
\begin{align}
\label{contfrac_1}
{\tilde A}(x)=\frac{a_1}{\displaystyle 1+\frac{\displaystyle
    a_2x}{\displaystyle 1+\frac{\displaystyle a_3x}{\displaystyle
      1+\ldots}}}\,,\\
\nonumber \\
\label{contfrac_2}
{\tilde B}(x)=\frac{b_1}{\displaystyle 1+\frac{\displaystyle
    b_2x}{\displaystyle 1+\frac{\displaystyle b_3x}{\displaystyle
      1+\ldots}}}\,,
\end{align}
\end{subequations}
where $a_1, a_2, a_3\ldots$ and $b_1, b_2, b_3\ldots$ are dimensionless
constants to be constrained, for instance, from observations of phenomena
near the event horizon. A few properties of the expansions
(\ref{contfrac})  are worth remarking. First, it
should be noted that \emph{at the horizon} only the first two terms of
the expansions survive, \ie
\begin{align}
{\tilde A}(0)={a_1}\,,
\qquad
{\tilde B}(0)={b_1}\,,
\end{align}
which in turn implies that near the horizon only the lowest-order terms
in the expansions are important. Conversely, \emph{at spatial infinity}
\begin{align}
{\tilde A}(1)=\frac{a_1}{\displaystyle 1+\frac{\displaystyle
    a_2}{\displaystyle 1+\frac{\displaystyle a_3}{\displaystyle
      1+\ldots}}}\,,
\qquad
{\tilde B}(1)=\frac{b_1}{\displaystyle 1+\frac{\displaystyle
    b_2}{\displaystyle 1+\frac{\displaystyle b_3}{\displaystyle
      1+\ldots}}}\,.
\end{align}
Finally, while the expansions (\ref{contfrac}) effectively contain an
infinite number of undetermined coefficients, we will necessarily
consider only the first $n$ terms. In this case, we simply need to set to
zero the $n$-th terms, since if $a_n=0=b_n$, then all terms of order
$m>n$ are not defined.

In practice, and as we will show in the rest of the paper, the superior
convergence properties of the continued fractions (\ref{contfrac}) are
such that the approximate metric they yield can reproduce all known (to
us) spherically symmetric metrics to arbitrary accuracy and with a
smaller set of coefficients. The inclusion of higher-order terms
obviously improves the accuracy of the approximation but in general
expansions truncated at $n=4$ are more than sufficient to yield the
accuracy that can be probed by present and near-future astronomical
observations.

\section{Comparison with the Johannsen-Psaltis parametrization}
\label{sec:JP}

To test the effectiveness of our approach in reproducing other known
spherically symmetric metric theories of gravity, we obviously start from
the Johannsen-Psaltis (JP) metric in the absence of
rotation~\cite{Johannsen:2011dh}. In this case, the black-hole line
element is spherically symmetric and is given by the following expression
for the slowly rotating dilaton black-hole solution
\begin{eqnarray}\label{JPbh}
ds^2&=&-\left[1+h(r)\right]\left(1-\frac{2\tilde{M}}{r}\right)dt^2 \nonumber\\
&& +
\left[{1+h(r)}\right]\left(1-\frac{2\tilde{M}}{r}\right)^{-1} dr^2 +
r^2 d\Omega^2\,,
\end{eqnarray}
where the function $h$ is a simple polynomial expansion in terms of the
expansion parameter $\tilde{M}/r$, \ie
\begin{equation}
\label{JP_expansion}
h(r) \equiv \sum_{n=1}^{\infty}\epsilon_n\left(\frac{\tilde{M}}{r}\right)^n
=\epsilon_1\frac{\tilde{M}}{r} +\epsilon_2\frac{\tilde{M}^2}{r^2} +
\epsilon_3\frac{\tilde{M}^3}{r^3} + \ldots\,.
\end{equation}

By construction, therefore, in the Johannsen-Psaltis metric the horizon
is located at
\begin{equation}
r_0=2\tilde{M}\,,
\end{equation}
while the relation between the ADM mass and the horizon mass $\tilde{M}$
is simply given by
\begin{equation}
M=\tilde{M}\left(1-\frac{\epsilon_1}{2}\right)\,.
\end{equation}

We can now match the asymptotic expansions for the metrics (\ref{ssbh})
and (\ref{JPbh}). More specifically, we can compare
Eqs.~\eqref{asympfix_1} and \eqref{JP_expansion}, to find that at
$\mathcal{O}\left((1-x)^2\right)$ the following relations apply between
our coefficients and those in the JP metric
\begin{subequations}
\label{JPrel}
\begin{align}
\epsilon&=-\frac{\epsilon_1}{2}\,, \\
a_0&=-\frac{1}{2}\left(\epsilon_1 - \frac{\epsilon_2}{2}\right)\,.
\end{align}
Similarly, comparing Eqs.~\eqref{asympfix_2} and \eqref{JP_expansion}, we
find that at $\mathcal{O}\left(1-x\right)$
\begin{equation}
b_0=\frac{\epsilon_1}{2}\,.
\end{equation}
\end{subequations}
It follows that if we set $\epsilon_1=\epsilon_2=0$, as done originally
in Ref.~\cite{Johannsen:2011dh}, then $\epsilon=0=a_0=b_0$, thus implying
that the PPN term are taken to be $\beta=\gamma=1$. We will not make this
assumption hereafter.

We can also match the expansions for the metrics (\ref{ssbh}) and
(\ref{JPbh}) near the horizon. More specifically, we can obtain algebraic
relations between our coefficients $a_n, b_n$ and the coefficients
$\epsilon_n$ of the JP metric by matching the $g_{tt}$ and $g_{rr}$
metric functions and their derivatives for $r \simeq r_0$ or $x \simeq
0$. A bit of tedious but straightforward algebra then leads to the
following expressions
\begin{subequations}
\label{as_bs_JP}
\begin{widetext}
\begin{align}
&a_1=h(r)\acond=\sum_{n=3}^{\infty}\frac{\epsilon_n}{2^n}=\frac{\epsilon_3}{8}+\frac{\epsilon_4}{16}+\frac{\epsilon_5}{32}\ldots\,, &
&\hskip -4.0cm b_1=h(r)\bcond=\sum_{n=2}^{\infty}\frac{\epsilon_n}{2^n}=\frac{\epsilon_2}{4}+\frac{\epsilon_3}{8}+\frac{\epsilon_4}{16}+\frac{\epsilon_5}{32}\ldots\,,
\\
\label{a2_b2_JP}
&a_2=-\frac{(r^3h)^\prime}{r^2h}\acond=\frac{\displaystyle\sum_{n=4}^{\infty}\frac{\epsilon_n(n-3)}{2^n}}{\displaystyle\sum_{n=3}^{\infty}\frac{\epsilon_n}{2^n}}\,,  &
&\hskip -4.0cm b_2=-\frac{(r^2h)^\prime}{rh}\bcond=\frac{\displaystyle\sum_{n=3}^{\infty}\frac{\epsilon_n(n-2)}{2^n}}{\displaystyle\sum_{n=2}^{\infty}\frac{\epsilon_n}{2^n}}\,,  \\
\label{a3_b3_JP}
&a_3=\frac{h(r_0)^2r^3}{2a_1a_2}\left(\frac{(r^3h)^\prime}{r^4h^2}\right)^\prime\acond\,, &
&\hskip -4.0cm b_3=\frac{h(r_0)^2r^2}{2b_1b_2}\left(\frac{(r^2h)^\prime}{r^2h^2}\right)^\prime\bcond\,,\\
\label{a4_b4_JP}
&a_4=\frac{1}{12a_1a_2a_3r^4}\left(r^2\left(r^2(r^3h)^\prime\right)^\prime\right)^\prime\acond
-\frac{(r^3h)^\prime}{4a_1a_2a_3r^2}\left(\frac{\left(r^2(r^3h)^\prime\right)^\prime}{(r^3h)^\prime}\right)^\prime\acond,  \\
&b_4=\frac{1}{12b_1b_2b_3r^3}\left(r^2\left(r^2(r^2h)^\prime\right)^\prime\right)^\prime\bcond
-\frac{(r^2h)^\prime}{4a_1a_2a_3r}\left(\frac{\left(r^2(r^2h)^\prime\right)^\prime}{(r^2h)^\prime}\right)^\prime\bcond\,,\\
&a_5 = \ldots\,,
\end{align}
\end{widetext}
\end{subequations}
where we have indicated with a prime $~^{\prime}$ the radial
derivative. Clearly, expressions~\eqref{as_bs_JP} can be easily extended
to higher orders if necessary.

A few remarks are worth doing. First, because of cancellations, the terms
$a_1,a_2,a_3\ldots$ do not depend on $\epsilon_1$ and $\epsilon_2$;
similarly, the terms $b_1,b_2,b_3\ldots$ do not depend on $\epsilon_1$,
but they do depend on $\epsilon_2$. Second, in the simplest case and the
one considered in Ref.~\cite{Johannsen:2011dh}, \ie when only
$\epsilon_3\neq0$, the coefficient $a_2$ vanishes and our approximant for
the function $N$ reproduces it exactly. Finally, and more importantly,
expressions \eqref{as_bs_JP} clearly show the rapid-convergence
properties of the expansions~\eqref{contfrac}. It is in fact remarkable
that a few coefficients only are sufficient to capture the infinite
series of coefficients needed instead in the JP approach [\cf for
  instance, expressions \eqref{a2_b2_JP} for the coefficients $a_1$ and
  $b_1$].

\section{Parametrization for dilaton black holes}
\label{sec:dilaton}

As another test of the convergence properties of our metric
parametrization we next consider a dilaton-axion black
hole~\cite{Garcia:1995}. When both the axion field and the spin vanish,
such a black hole is described by a spherically symmetric metric with
line element
\begin{eqnarray}
\label{dbh}
ds^2=&&-\left(\frac{\rho-2\mu}{\rho+2b}\right)dt^2
+\left(\frac{\rho+2b}{\rho-2\mu}\right)d\rho^2
+(\rho^2+2b\rho)d\Omega^2\,. \nonumber\\
\end{eqnarray}
The radial coordinate $r$ and the ADM mass $M$ are expressed, respectively,
as
\begin{equation}
r^2=\rho^2+2b\rho,\qquad \qquad M=\mu+b\,,
\end{equation}
where $b$ is the dilaton parameter.

By comparing now the expansions of (\ref{ssbh}) and (\ref{dbh}) at
spatial infinity, we find that
\begin{subequations}
\label{dilaton0}
\begin{align}
\epsilon & =\sqrt{1+\frac{b}{\mu}}-1\,, \\
a_0 & =\frac{b}{2\mu}\,, \\
b_0 & =0 \,,
\end{align}
\end{subequations}
Similarly, by comparing the near-horizon expansions we find the other
coefficients, which also depend on $b/\mu$ only and are given by
\begin{subequations}
\label{dilaton1}
\begin{align}
&a_1=2\sqrt{1+\frac{b}{\mu}}+\frac{1}{1+{b}/{(2\mu)}}-
3-\frac{b}{2\mu}\,,\\
&b_1=\frac{\sqrt{1+{b}/{\mu}}}{1+{b}/({2\mu})}-1\,,\\
&a_2=\frac{\displaystyle\sqrt{1+\frac{b}{\mu}} - \frac{1}{2}  -
\frac{b^2}{4\mu^2}+\frac{b}{2\mu}
\left(\sqrt{1+\frac{b}{\mu}}-1\right)}
{\displaystyle\left(1+\frac{b}{2\mu}\right)^2}\,,\\
&b_2=\frac{\sqrt{1+{b}/{\mu}}}{1+{b}/({2\mu})}-1-\frac{b^2}{(b+2\mu)^2}\,.
\end{align}
\end{subequations}

\begin{figure*}
\resizebox{\linewidth}{!}
{\includegraphics*{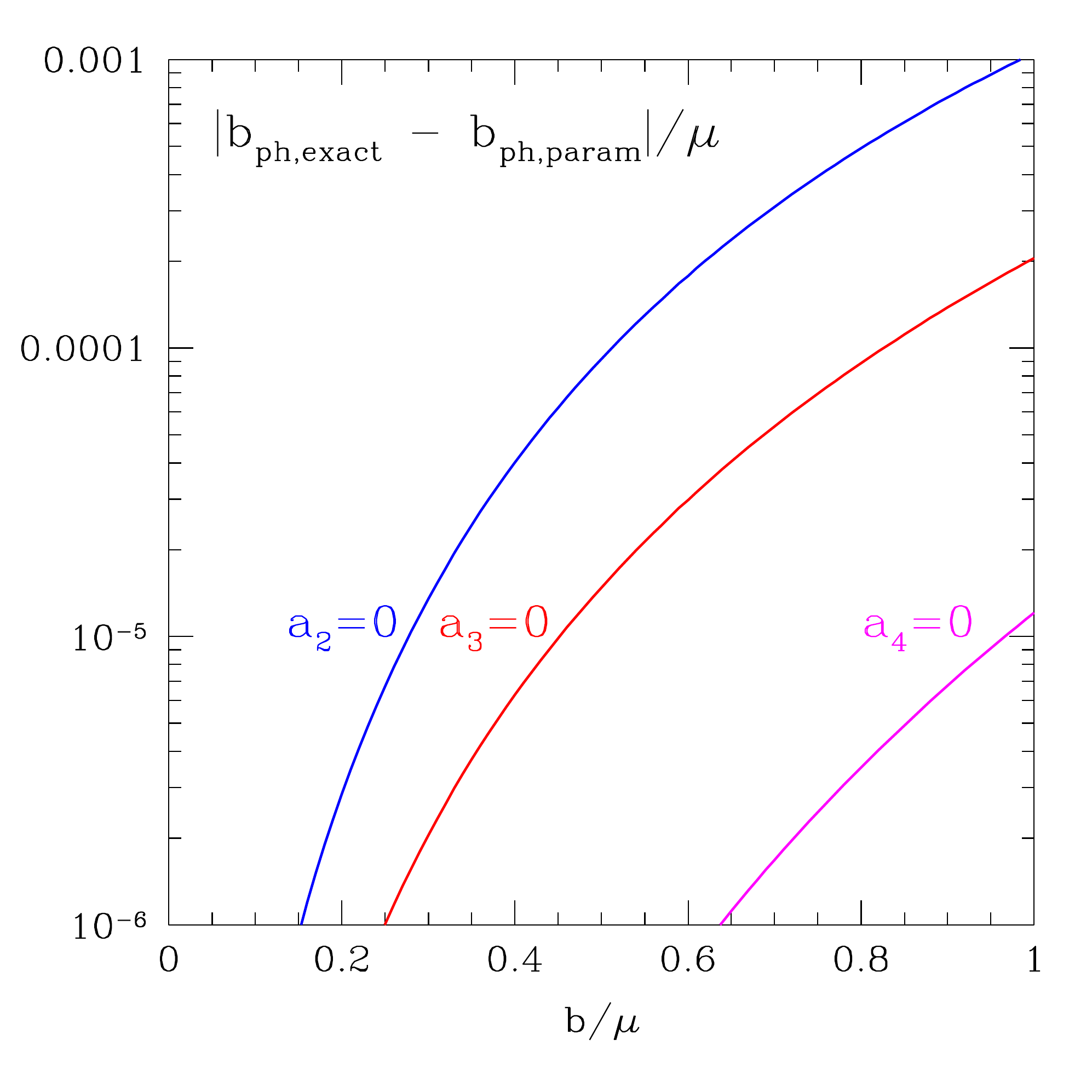}
  \includegraphics*{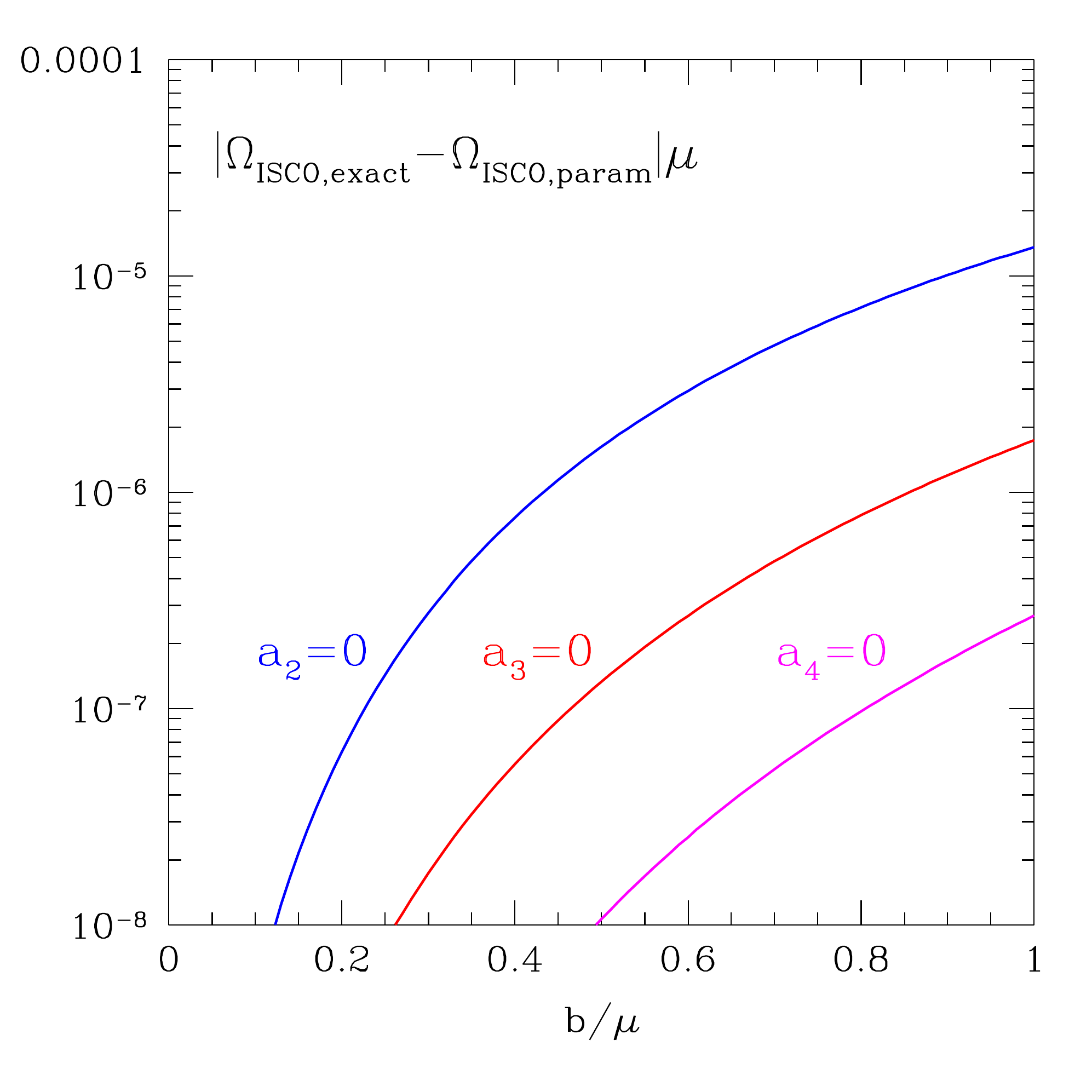}}
\caption{\textit{Left panel:} Difference between the exact values of the
  dilaton black hole orbit impact parameter for a circular orbit $b_{\rm
    ph}$ and the values obtained using the continued-fraction expansions
  \protect(\ref{contfrac}). The results are shown as a function of the
  dimensionless strength of the dilaton parameter $b/\mu$. Different
  lines refer to different levels of approximation, \ie $a_2=0$ (blue
  line), $a_3=0$ (red line), and $a_4=0$ (magenta line). Note that even
  when $a_2=0$, the differences are $\lesssim 10^{-4}$ for $b\lesssim
  \tfrac{1}{2}\mu$. \textit{Right panel:} The same as in the left panel
  but for the ISCO frequency.}\label{fig:freqdiff}
\end{figure*}

It is clear that $a_1$ and $b_1$ vanish if $b=0$, in which case we
reproduce the line element of the Schwarzschild black hole exactly. If
$b>0$, on the other hand, we could in principle calculate as many
coefficients of the continued fractions~\eqref{contfrac} as needed; in
practice already the very first ones suffice. For example, for $b/\mu=1$
and setting $a_3=0$, the maximum relative difference between the exact
and the expanded expression for the metric function $g_{tt}$ is $\lesssim
3\times 10^{-4}$. This relative difference becomes $\lesssim 3\times
10^{-6}$ if the order is increased of one, \ie if $a_4=0$ (see also the
discussion below on Fig.~\ref{fig:freqdiff}).

\section{Observable quantities within the parametrization framework}
\label{sec:observe}

A high precision in the mapping of the metric functions does not
necessarily translate in an equivalent accurate measure of near-horizon
phenomena. Hence, to further test the reliability of our
continued-fraction expansions (\ref{contfrac}), we next compare a number
of potentially observable quantities for a spherically symmetric dilaton
black hole and for a black hole in Einstein-aether theory, respectively.
More specifically, we calculate: the impact parameter for the photon
circular orbit, the orbital frequency for the innermost stable circular
orbit, and the quasinormal ringing of a massless scalar field. For all of
these quantities, the metric is either expressed analytically [\ie
  Eq. \eqref{dbh} for a dilation black hole] or numerically [\ie for a
  black hole in Einstein-aether theory], or in its parametrized form
[\ie via the coefficients~\eqref{dilaton0}--\eqref{dilaton1} for a
  dilation black hole].

\subsection{Photon circular orbit and the innermost stable circular orbit}

In a spherically symmetric spacetime, a photon circular orbit is defined
as the null geodesic at radial position $r=r_{\rm ph}$ for which the
following equations are satisfied
\begin{align}
ds^2&=-N(r_{\rm ph})^2dt^2+r_{\rm ph}^2d\phi^2=0\,,\\
d^2r_{\rm ph}&=-\frac{N'(r_{\rm ph})N(r_{\rm ph})^3}{B(r_{\rm ph})^2}dt^2+
\frac{N(r_{\rm ph})^2r_{\rm ph}}{B(r_{\rm ph})^2}d\phi^2=0\,,\nonumber\\
\end{align}
where we have implicitly assumed $\theta=\pi/2$ because of the absence of
a preferred direction. From these equations we find that the equation for
the radius is given by
\begin{equation}\label{photon}
r_{\rm ph}=\frac{N(r_{\rm ph})}{N'(r_{\rm ph})}\,,
\end{equation}
and that the corresponding orbital frequency $\Omega_{\rm ph}$ is
\begin{equation}
\label{photonfreq}
\Omega_{\rm ph}=
\left.\frac{d\phi}{dt}\right\vert_{r=r_{\rm ph}}=
\sqrt{\frac{N'(r_{\rm ph})N(r_{\rm ph})}{r_{\rm ph}}}=
\frac{N(r_{\rm ph})}{r_{\rm ph}}\,.
\end{equation}
Note that expression~\eqref{photonfreq} depends only on the coefficients
$\epsilon$ and $a_n$, but not on the $b_n$ coefficients [\cf
  Eq.~\eqref{N2}]. We then define the impact parameter of the photon
circular orbit (not to be confused with the dilaton parameter) as
\begin{equation}
b_{\rm ph}=\frac{1}{\Omega_{\rm ph}}\,.
\end{equation}
whose analytic expression in the case of a dilaton black hole
is~\cite{Wei:2013kza}
\begin{equation}
\label{bph_exact}
b_{\rm ph}=
\mu\sqrt{\frac{27+36{b}/{\mu}+8{b^2}/{\mu^2}+
\left(9+8{b}/{\mu}\right)^{3/2}}{2}}
\,,
\end{equation}
which reduces to $b_{\rm ph}/\mu = 3 \sqrt{3}$ in the case of a
Schwarzschild black hole.

In the left panel of Fig.~\ref{fig:freqdiff} we show the difference
between the exact values of $b_{\rm ph}$ computed via
Eq.~\eqref{bph_exact} and the ones obtained after solving numerically
Eq.~(\ref{photon}) and making use the continued-fraction expansions
(\ref{contfrac}) with
coefficients~\eqref{dilaton0}--\eqref{dilaton1}. The differences are
shown as a function of the dimensionless dilaton parameter $b/\mu$ and
different lines refer to different levels of approximation, \ie when
setting $a_2=0$ (blue line), $a_3=0$ (red line), and $a_4=0$ (magenta
line). The figure shows rather clearly that already when setting $a_2=0$,
that is, when retaining only the coefficients $\epsilon, a_0, b_0, a_1, $
and $b_1$, the differences in the impact parameter are of the order of
$\sim 10^{-4}\,\mu$ for $b \sim \tfrac{1}{2} \mu$. These differences
become larger with larger dilaton parameter, but when $a_4=0$ they can
nevertheless be reduced to be $\sim 10^{-6}\,\mu$ even for $b \sim \mu$.

In a similar way, we can calculate the innermost stable circular orbit
(ISCO) exploiting the fact that the geodesic motion of a massive particle
in the equatorial plane can be reduced to the one-dimensional motion
within an effective potential
\begin{equation}\label{potential}
V_{\rm eff}(r)=\frac{E^2}{N^2(r)}-\frac{L^2}{r^2}-1 \,,
\end{equation}
where $E$ and $L$ are the constants of motion, \ie energy and angular
momentum, respectively. A circular orbit then is the one satisfying the
following conditions
\begin{equation}\label{circular}
V_{\rm eff}(r)=0=V_{\rm eff}^\prime(r)\,.
\end{equation}
while the ISCO is defined as the radial position $r_o$ at which
\begin{equation}\label{ISCO}
V_{\rm eff}^{\prime\prime}(r_o)=0\,.
\end{equation}

Substituting (\ref{potential}) into (\ref{circular}) and (\ref{ISCO}), we
obtain the following algebraic equation for the ISCO radius $r_o$,
\begin{equation}
\label{ISCOeq}
3N(r_o)N^\prime(r_o)-3r_oN^{\prime\ 2}(r_o)+r_oN(r_o)N^{\prime\prime}(r_o)=0,
\end{equation}
which we can solve numerically to calculate the corresponding orbital
frequency as
\begin{equation}
\label{O_ISCO_par}
\Omega_{_{\rm ISCO}}=\sqrt{\frac{N'(r_o)N(r_o)}{r_o}}\,.
\end{equation}
Here too, expression~\eqref{O_ISCO_par} depends only on the coefficients
$\epsilon$ and $a_n$, but not on the $b_n$ coefficients.

The ISCO frequency~\eqref{O_ISCO_par} can of course be compared with the
exact expression in the case of a dilaton black hole, which is given
by
\begin{equation}
\label{O_ISCO_ex}
\Omega_{_{\rm ISCO}} = \frac{1}{2\mu}
\sqrt{\frac{\kappa}{(1+\kappa)(1+\kappa+\kappa^2)^3}}\,,
\end{equation}
where
\begin{equation}
\kappa \equiv\left(1+\frac{b}{\mu}\right)^{1/3}\,,
\end{equation}
and expression~\eqref{O_ISCO_ex} reduces to the well-known result of
$\Omega_{_{\rm ISCO}}\,M = (1/6)^{3/2}$ in the case of a Schwarzschild
spacetime.

\begin{figure*}
\resizebox{\linewidth}{!}
{\includegraphics*{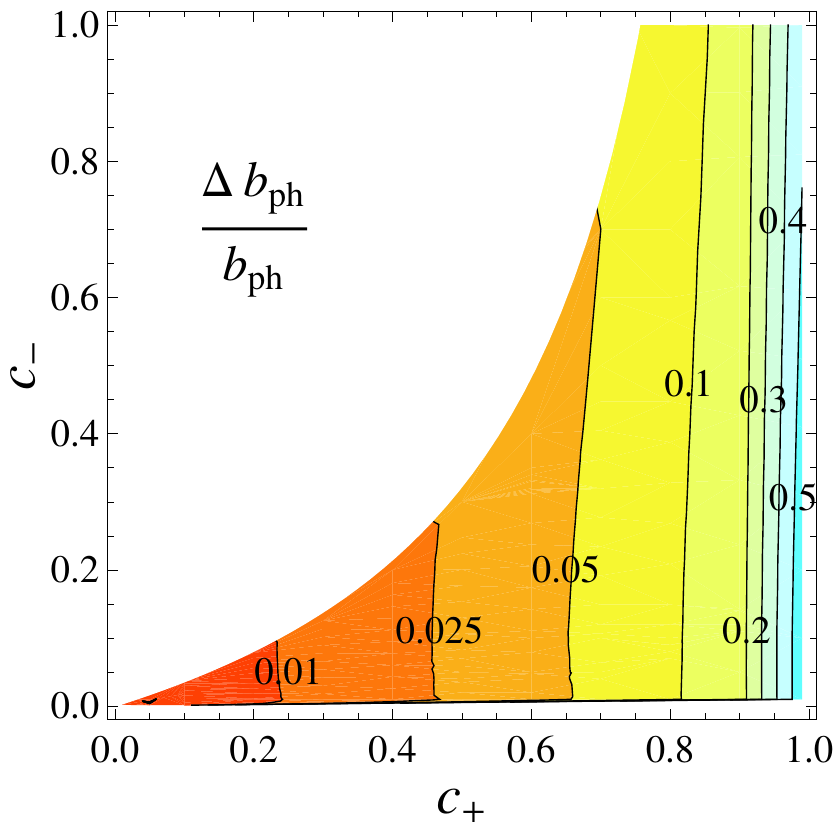}
\includegraphics*{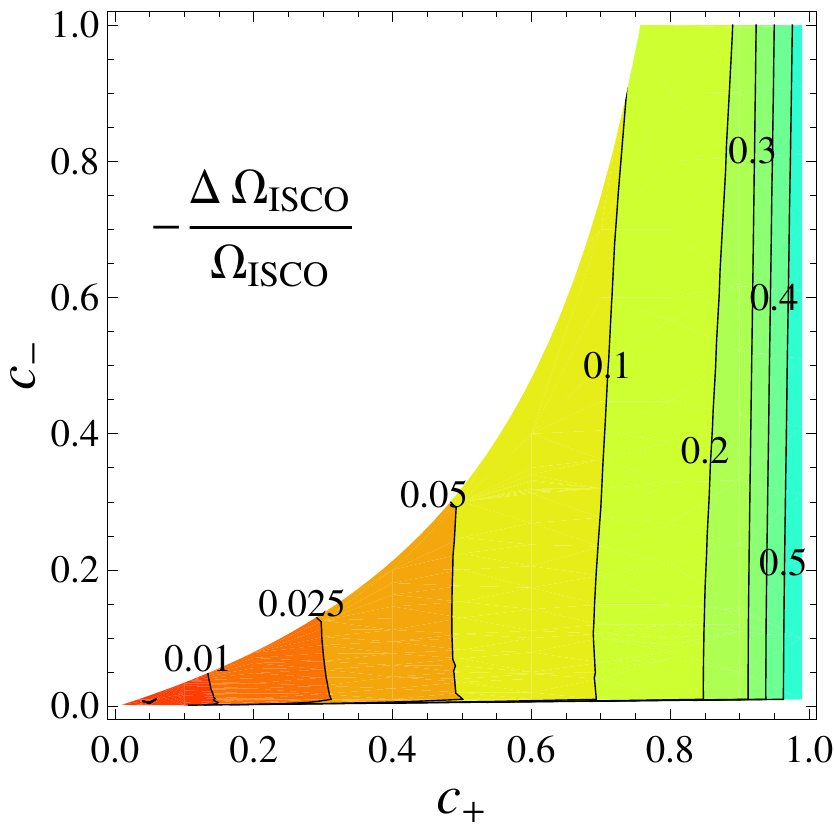}}
\caption{\textit{Left panel:} Relative difference in the photon circular
  orbit impact parameter $b_{\rm ph}$ between general relativity and the
  alternative Einstein-aether theory (\cf Fig.~4 of
  Ref.~\cite{Barausse:2011pu}).  The differences are reported within the
  mathematically allowed ranges for the aether parameters $c_+$ and
  $c_-$. The contours correspond to the following values (from left to
  right): $0.01$, $0.025$, $0.05$, $0.1$, $0.2$, $0.3$, $0.4$,
  $0.5$. \textit{Right panel:} the same as in the left panel but for the
  ISCO frequency (\cf Fig.~2 of Ref.~\cite{Barausse:2011pu}). The impact
  parameter and ISCO frequencies were calculated using continued-fraction
  expansions with $a_3=0$.}\label{fig:AE}
\end{figure*}

A comparison between the values of the ISCO frequency estimated from
expressions~\eqref{O_ISCO_par} and~\eqref{O_ISCO_ex} is shown in the
right panel of Fig.~\ref{fig:freqdiff}, which reports the differences in
units of $\mu$ and as a function of the dimensionless dilaton parameter
$b/\mu$. As for the left panel, different curves refer to different
levels of approximation, \ie $a_2=0$ (blue line), $a_3=0$ (red line), and
$a_4=0$ (magenta line). Also in this case, the differences in the ISCO
frequency are of the order of $\sim 10^{-6}\,\mu$ for $b \sim
\tfrac{1}{2} \mu$ and can reduced to be $\sim 10^{-7}\,\mu$ even for $b
\sim \mu$ by including higher-order coefficients.

Finally, we have compared the values for the impact parameter and the
ISCO frequency also for another spherically symmetric black hole, namely,
the one appearing in the alternative Einstein-aether theory of gravity
\cite{JacobsonAE}. In this case, the metric is not known analytically,
but we have used the numerical data for the metric functions as discussed
in Ref.~\cite{Barausse:2011pu}. More specifically, for a large number of
pairs of the aether parameters $c_+$ and $c_-$, we have obtained a
numerical approximation of the $g_{tt}$ metric function $N(r)$ in terms
of the coefficients $\epsilon$, $a_0$, $a_1$, and $a_2$ of our
continued-fraction expansions~\eqref{contfrac}. Using these coefficients,
we have then calculated numerically the values of $b_{\rm ph}$ and
$\Omega_{_{\rm ISCO}}$ as discussed above and compared with the
corresponding values in general relativity.

The results of this comparison are reported in Fig.~\ref{fig:AE}, whose
left panel refers to the impact parameter for a circular photon orbit,
while the right panel to the ISCO frequency. The two panels are meant to
reproduce Figs. 2 and 4 of Ref.~\cite{Barausse:2011pu} and they do so
with an accuracy of fractions of a percent. Note that the differences in
$b_{\rm ph}$ and $\Omega_{_{\rm ISCO}}$ with respect to general
relativity can be quite large for certain regions of the space of
parameters (\eg $c_+ \simeq 1$). These regions, however, are de-facto
excluded by the observational constraints set by binary pulsars (see the
discussion in Ref.~\cite{Yagietal_13}).

\subsection{Quasinormal ringing}

Another way to probe whether the metric parametrization~\eqref{ssbh} and
the continued fraction expansions~\eqref{contfrac} represent an effective
way to reproduce strong-field observables near a black hole is to compare
the response to perturbations. We recall, in fact, that if perturbed, a
black hole will start oscillating. Such oscillations, commonly referred
to as ``quasinormal modes'', represent exponentially damped oscillations
that, at least at linear order, do not depend on the details of the
source of perturbations, but only on the black hole parameters (see
\cite{Konoplya:2011qq} for a review). The relevance of these oscillations
is that they probe regions of the spacetime that are close to the light
ring, but are global and hence do not depend on a single radial
position. At the same time, the gravitational-wave signal from a
perturbed black hole can be separated from a broad class of the
environmental effects, allowing us to expect a good accuracy of the
quasinormal modes' measurement \cite{Barausse:2014tra}. Furthermore, both
of the continued-fraction expansions~\eqref{contfrac} are involved and
hence also some of the $b_n$ coefficients will be nonzero.

\begin{figure*}
\resizebox{\linewidth}{!}{
\includegraphics*{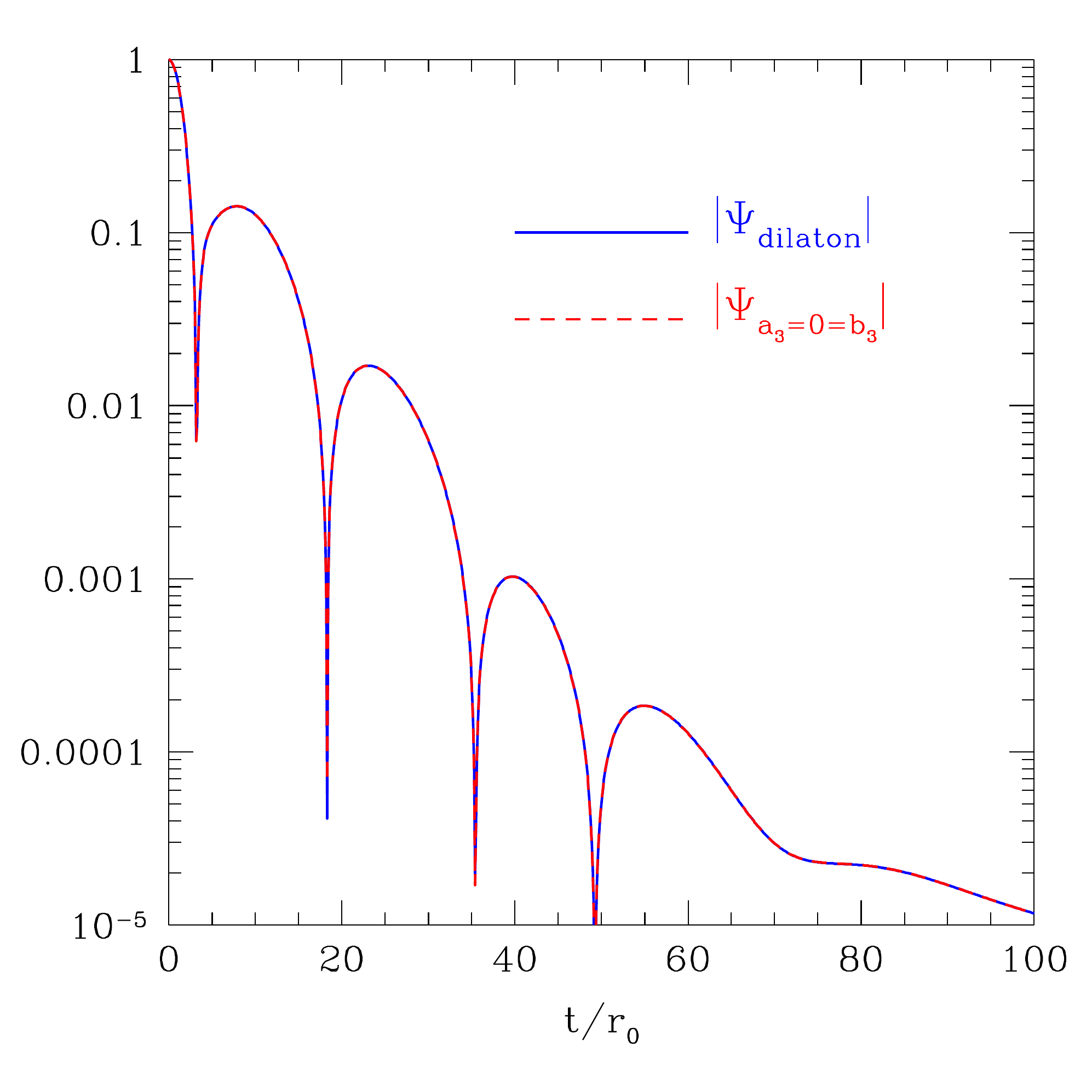}
  \includegraphics*{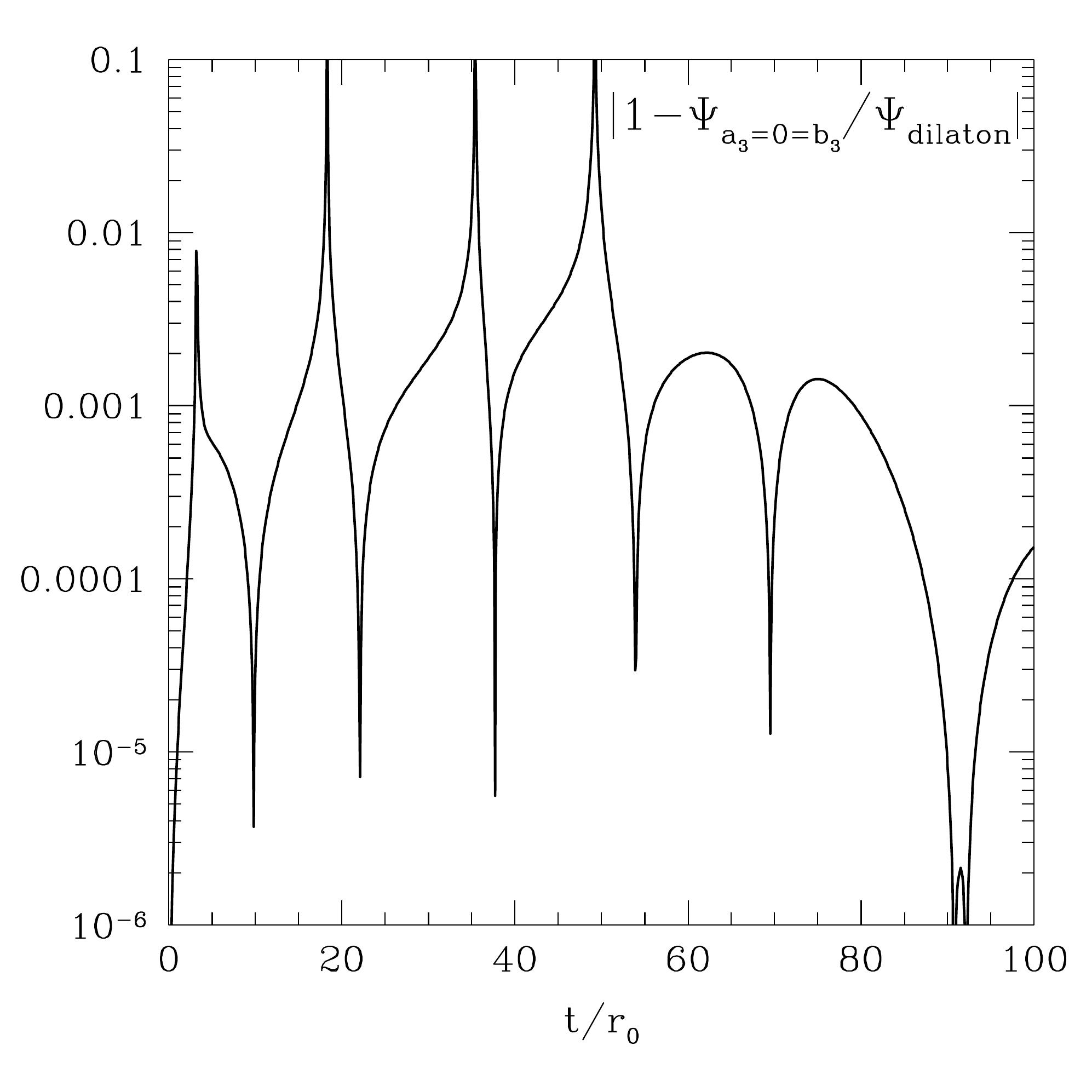}}
\caption{\textit{Left panel:} Evolution of the scattered scalar field
  $|\Psi|$ for the $\ell=0$ perturbations at $r=2r_0$ as computed using
  the exact dilaton black hole metric with $b/\mu=1$ (blue solid line) or
  the corresponding parametrized form with $a_3=0=b_3$ (red dashed
  line). \textit{Right panel:} relative difference in the evolution of
  $|\Psi|$ shown in the left panel.}
\label{fig:timedomain}
\end{figure*}

For simplicity, we have considered the evolution of a massless scalar
field $\Phi$ as governed by the Klein-Gordon equation
\begin{equation}
\label{boxphi}
\square\Phi=0\,,
\end{equation}
where $\square$ is the Dalambertian operator. Substituting
in~\eqref{boxphi} the ansatz
\begin{equation}
\Phi(t,r,\theta,\phi)=\Psi(t,r)Y_{\ell}(\theta,\phi)/r,
\end{equation}
where $Y_{\ell}(\theta,\phi)$ are Laplace's spherical harmonics, we
obtain for each multipole number $\ell$ the following wave-like equation,
\begin{equation}
\left(\frac{\partial^2}{\partial t^2}-\frac{\partial^2}{\partial r_*^2}+
V_{\ell}(r)\right)\Psi(t,r) = 0\,,
\end{equation}
where we have introduced the (tortoise-like) radial coordinate
\begin{equation}
dr_*=\frac{B(r)}{N^2(r)}dr\,,
\end{equation}
and the effective potential is given by
\begin{equation}
V_{\ell}(r) = \frac{\ell(\ell+1)}{r^2} N^2(r) +
\frac{1}{r}\frac{d}{dr_*}\frac{N^2(r)}{B(r)}\,.
\end{equation}

It was shown in Ref.~\cite{Aether} that the rational approximation for
$N(r)$ and $B(r)$ in some region near the black hole horizon in reduced
Einstein-aether theory allows one to calculate accurately at least the
quasinormal modes with the longest damping time. In order to test our
approximation in the case of dilaton black hole, we have compared the
black hole response in the time domain, found using either the exact
representation of the metric~\eqref{dbh} or the parametrized one via the
coefficients~\eqref{dilaton0}--\eqref{dilaton1}.

The numerical solution of the evolution equation was made using a
characteristic integration method that involves the light-cone variables
$u \equiv t - r_*$ and $v \equiv t + r_*$~\cite{Gundlach:1993tp}, with
initial data specified on the two null surfaces $u = u_{0}$ and $v =
v_{0}$. The results of these calculations are shown in
Fig.~\ref{fig:timedomain}, whose left panel reports the $\ell=0$ solution
of the scalar field at $r=2r_0$ as function of time both in the case of
an exact dilaton black hole (blue line) and of the corresponding
parametrized expansion (red line). The relative differences are clearly
very small already with $a_3=0=b_3$, as shown in the right panel of
Fig.~\ref{fig:timedomain}, and amounting at most to fractions of a
percent.

\section{Slowly rotating black holes}\label{sec:rotation}

We are in position now to make the first step towards the parametrization
of black holes that are not spherically symmetric. We believe that the
natural way to choose the parameters is taking into account the
asymptotical behaviour of the corresponding metric, which is defined by
multipole moments \cite{Ryan:1995-1997}, as well as its near-horizon
behaviour. Unfortunately, only a very limited number of such metrics is
known in alternative theories of gravity that can be used for
comparison. Indeed, to the best of our knowledge, a black hole with
independent multipole moments was studied only in general relativity and
discussed in Refs.~\cite{Collins:2004ex,Glampedakis:2005cf}. At the same
time, even the parametrization of an axisymmetric stationary black hole
is far from being a trivial question, since the corresponding metric is
defined by four functions of two variables.

As a warm-up exercise, in this section we will consider spacetime metrics
having only a small deviation from the spherical symmetry and hence
extend the general expression~\eqref{ssbh} by introducing a new function
$\omega$ in the $g_{t\phi}$ metric function and by retaining it only at
the first order, \ie
\begin{eqnarray}
\label{srbh}
ds^2&=&-N^2(r)dt^2+\frac{B^2(r)}{N^2(r)}dr^2
+r^2 d\Omega^2
\nonumber\\
&&-2\omega(r,\theta) r^2\sin^2\theta \, dtd\phi + \mathcal{O}(\omega^2)\,,
\end{eqnarray}
and with the condition that $\omega$ has a falloff with radius that is
faster than $r^{-1}$, \ie that
\begin{equation}\label{reqaxion}
r\omega(r,\theta)\ll 1\,,
\end{equation}
implying that the event horizon remains at $r=r_0$~\footnote{Determining
  in the metric~\eqref{srbh} the location where $g^{rr}=0$ will also
  involve the square of the metric function $g_{t\phi}$, which we take to
  be zero in the slow-rotation approximation.}.

Because we are not considering any consistent (alternative) theory of
gravity, but we are simply prescribing ad-hoc expression for the metric,
we cannot impose additional constraints on the function
$\omega$. However, if we assume that the function $\omega$ depends on the
radial coordinate $r$ only, then we obtain a metric which can be
associated with a slowly rotating black hole in Ho\v{r}ava-Lifshitz
theory \cite{Barausse:2012qh}, in Einstein-aether gravity
\cite{Barausse:2013nwa}, in Chern-Simons modified gravity
\cite{Konno:2009kg}, or with dilatonic Einstein-Gauss-Bonnet
\cite{Pani:2009wy} and dilaton-axion black holes
\cite{Garcia:1995}.

In this case, the asymptotic behavior is given
\begin{align}
\omega(r,\theta) \to \omega(r) &= \frac{2J}{r^3}+{\cal O}\left(r^{-4}\right)
\nonumber \\
&= \frac{2J}{r_0^3}(1-x)^3+{\cal O}\left((1-x)^4\right)\,,
\end{align}
where $J$ is the spin of the black hole and we take it to be $J\ll M^2$.

The parametrization of the function $\omega$ can then be made in analogy
with what was done for the nonrotating case and again we use a Pad\'e
approximation in terms of continued fractions in the form
\begin{align}
\label{dabhpar}
r\omega(r) & = \left(\frac{r_0}{1-x}\right)\omega(x)\nonumber \\
& = \omega_0 (1-x)^2+\frac{\omega_1(1-x)^3}{\displaystyle 1+
\frac{\displaystyle \omega_2x}{\displaystyle 1+
\frac{\displaystyle \omega_3x}{\displaystyle 1+\ldots}}}\,.
\end{align}
Since $r_0 \omega(x) = \omega_0 (1-x)^3 + {\cal O}\left((1-x)^4\right)$,
the first coefficient is simply given by $\omega_0 \equiv 2J/r_0^2$,
while the higher-order ones, $\omega_1,\omega_2,\omega_3\ldots$, are
fixed by comparing series expansion of $\omega$ near the event horizon
$r_0$.

As an example, we consider the first-order correction to the
dilaton black hole (\ref{dbh}) due to rotation given by the following
line element~\cite{Garcia:1995}
\begin{eqnarray}
\label{dabh}
ds^2&=&-\left(\frac{\rho-2\mu}{\rho+2b}\right)dt^2+
\left(\frac{\rho+2b}{\rho-2\mu}\right)d\rho^2 \nonumber \\
&&-
\left[\frac{4a(\mu+b)}{\rho+2b}\right]\sin^2\theta dt\,d\phi
+(\rho^2+2b\rho)d\Omega^2
\nonumber\\
&&+\ {\cal O}(a^2).
\end{eqnarray}
By comparing the asymptotical and near-horizon expansions we find that
\begin{subequations}
\label{o0o1}
\begin{align}
\omega_0&=\frac{a}{2\mu}\,,\\
\omega_1&=\frac{a}{2\mu}\left(\sqrt{\frac{\mu}{\mu+b}}-1\right)\,,\\
\omega_2&=\frac{\sqrt{\mu(\mu+b)}-\mu-b}{2\mu+b}\,,\\
\omega_3&=\frac{\mu b}{(2\mu+b)^2}\,.
\end{align}
\end{subequations}
Since we consider $a\ll\mu$, the coefficients (\ref{o0o1}) imply that
$\omega_0\ll1$ and $\omega_1\ll1$, thus satisfying the constraint
(\ref{reqaxion}). The other coefficients are not small and depend on the
dilaton parameter only. Of course, it is possible to find as many
coefficients in (\ref{dabhpar}) as needed for an accurate approximation
for the function $\omega$.

In order to test the convergence properties of (\ref{dabhpar}) we again
study the ISCO frequency for the equatorial orbits (\ie $\theta=\pi/2$)
of a massless particle in the background of a slowly rotating
dilaton black-hole metric (\ref{srbh}). In this case, the effective
potential reads
\begin{equation}
V_{\rm eff}(r)=
\frac{E^2 r^2-2EL\omega(r)r^2-L^2N^2(r)}{N(r)^2r^2+\omega(r)^2r^4}\,.
\end{equation}
We assume now that the energy $E$ and the angular momentum $L$ are
positive, thus implying that $a>0$ for the co-rotating and $a<0$ for the
counter-rotating particles, respectively. We then solve numerically the
set of equations
\begin{subequations}
\begin{align}
&V_{\rm eff}(r_o)=0\,,\\
&V_{\rm eff}'(r_o)=0\,,\\
&V_{\rm eff}''(r_o)=0\,,
\end{align}
\end{subequations}
finding at the radial coordinate of the ISCO $r_o>r_0$ the corresponding
frequency $\Omega_{_{\rm ISCO}}$
\begin{eqnarray}
\Omega_{_{\rm ISCO}}&=&\frac{-g_{t\phi}'+\sqrt{{g_{t\phi}'}^2-
g_{tt}'g_{\phi\phi}'}}{g_{\phi\phi}'}\Biggr|_{r=r_o}=
\omega(r_o)+\frac{\omega'(r_o)r_o}{2}\nonumber\\
&+&\sqrt{\frac{N(r_o)N'(r_o)}{r_o}+
\left(\omega(r_o)+\frac{\omega'(r_o)r_o}{2}\right)^2}\,.
\end{eqnarray}

\begin{figure}
\resizebox{\linewidth}{!}
{\includegraphics*{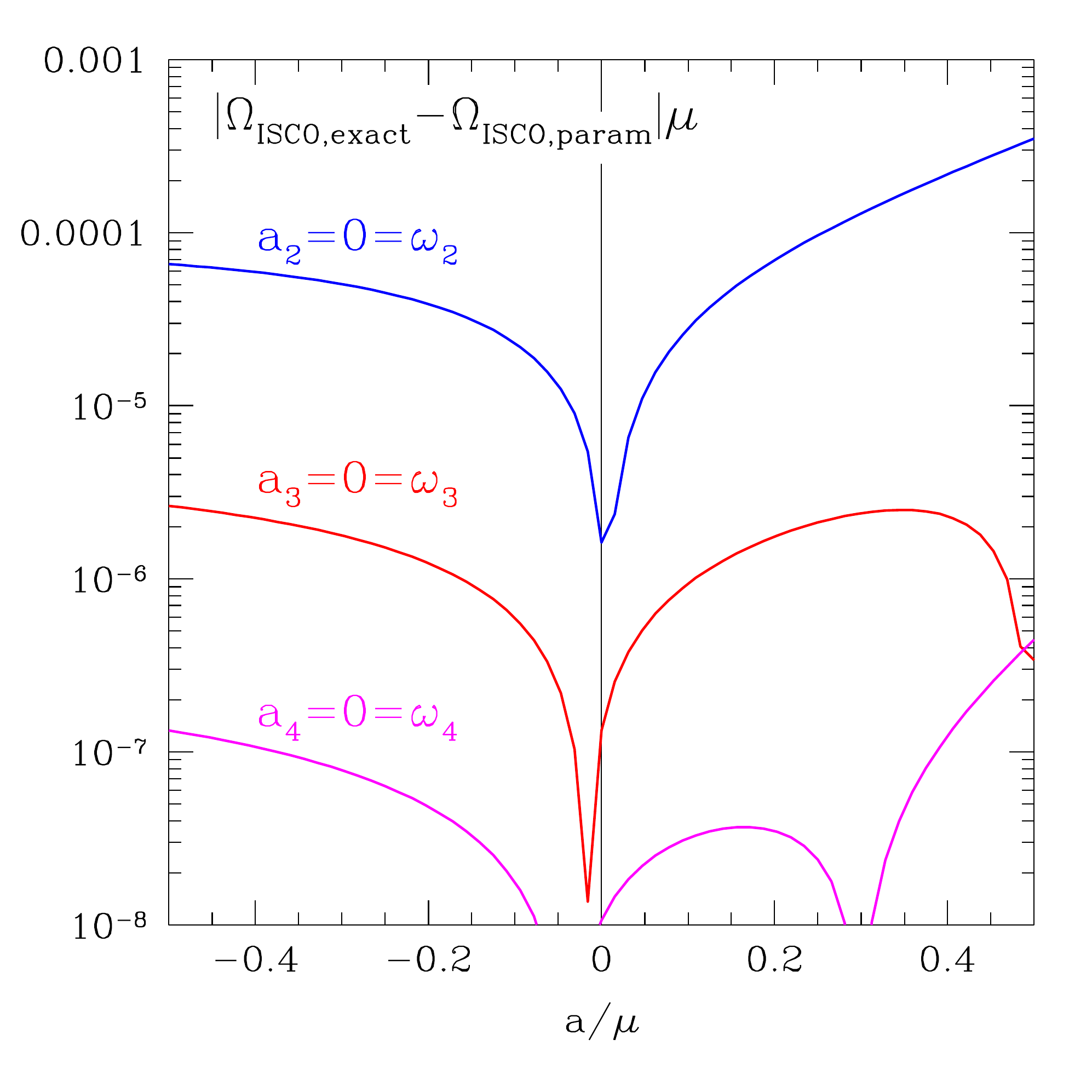}}
\caption{Difference between the exact values of the ISCO frequencies in
  the equatorial plane for the dilaton black hole in the
  slow-rotation approximation regime and the values obtained using the
  continued-fraction expansions. Different curves, shown as functions of
  the rotation parameter, refer to the different levels of approximation:
  $a_2=0=\omega_2$ (blue line), $a_3=0=\omega_3$ (red line), and
  $a_4=0=\omega_4$ (magenta line). In all cases we have taken a reference
  value of $b=\mu/2$.}\label{fig:ISCO-axion-dilaton}
\end{figure}

Of course, these frequencies can be computed also for the parametrized
metric (\ref{o0o1}) at different level of approximation. A comparison
between the two calculations is summarized in
Fig.~\ref{fig:ISCO-axion-dilaton}, which shows the absolute value of the
difference between the exact value of $\Omega_{_{\rm ISCO}}$ and the
approximate one as a function of the normalized spin parameter
$a/\mu$. As in the previous figures, here too different curves (all
computed for $b=\mu/2$) refer to different degrees of approximation:
$a_2=0=\omega_2$ (blue line), $a_3=0=\omega_3$ (red line), and
$a_3=0=\omega_3$ (magenta line). Also in this case it is apparent that
the use of a larger number of coefficients in continued fraction
expansions (\ref{contfrac_1}) and (\ref{dabhpar}), leads to a monotonic
increase of the accuracy of the ISCO frequency.

As a concluding remark we note that a possible and rather popular
approach to extend the parametrization~\eqref{ssbh} to rotating black
holes would be the application of the Newman-Janis algorithm~\cite{NJ65}
to the metric (\ref{ssbh}) after having fixed the parameters $\epsilon,
a_0, b_0, a_1, b_1,\ldots$. Although there is no proof that such a
rotating configuration corresponds to a black hole solution in the same
theory, the method works for some theories, \eg the application of the
Newman-Janis method to the metric (\ref{dbh}) allows one to obtain the
metric for the axion-dilaton black hole
\cite{Yazadjiev:1999ce}. Therefore, it would be interesting to compare
the coefficients (\ref{o0o1}) with those obtained at first order after
the application of the Newman-Janis algorithm.

Yet, it is clear that this approach would not provide us with the most
generic form for an axisymmetric black hole, simply because, in general,
the geometry cannot be parametrized by one rotation parameter only. What
is needed, instead, is a general framework which naturally comprises a
set of parameters that account for the multipole moments of the spacetime
and are not necessarily restricted to follow the relations expressed in
terms of mass and angular momentum that apply for a Kerr black
hole. Investigating this approach is beyond the scope of this initial
paper, but will be the focus of our future work.

\section{Conclusions}

We have proposed a new parametric framework to describe the spacetime of
spherically symmetric and slowly rotating black holes in generic metric
theories of gravity. The new framework provides therefore a link between
astronomical observations of near-horizon physics with the properties of
black holes in alternative theories of gravity, and which would predict
deviations from general relativity. Unlike similar previous attempts in
this direction, our approach is based on two novel choices. First, we
use a continued-fraction expansion rather than the traditional Taylor
expansion in powers of $M/r$, where $M$ and $r$ are respectively the mass
of the black hole and a generic radial
coordinate~\cite{Johannsen:2011dh,Cardoso:2014rha}.  Second, the
expansion is made in terms of a compactified radial coordinate with
values between zero and one between the horizon and spatial
infinity. These choices lead to superior convergence properties and
allows us to approximate a number of known metric theories with a much
smaller set of coefficients.

These parameters can be calculated very accurately for any chosen
spherically symmetric metric and, at the same time, they can be used via
astronomical observations to measure near-horizon phenomena, such as
photon orbits or the position ISCO. As a result, the new parametrization
provides us with powerful tool to efficiently constrain the parameters of
alternative theories using future astronomical observations.

Another important advantage of our approach is that we can use not only
the asymptotic parameters from the PPN expansion, but also the
near-horizon parameters, which are well-captured already by the first
lowest-order coefficients. More specifically, the most important
parameters for the near-horizon geometry are expressed simply in terms of
the coefficients $\epsilon$ (which relates the ADM mass and the event
horizon), $a_1$, $b_1$, and $\omega_1$. The use of other higher-order
parameters increases the accuracy of the approximation, but does not
change significantly the observable quantities. The latter, in fact, are
captured to the precision of typical near-future astronomical
observations already at the lowest order.

The rapid convergence of our expansion is also useful for the analysis of
black-hole spacetimes in alternative theories where the metric is known
only numerically. Using as a practical example the alternative
Einstein-aether theory of gravity, we have shown that it is possible to
reproduce to arbitrary accuracy the numerical results by using a small
set of coefficients in the continued-fraction expansion. In turn,
adopting such coefficients it is also possible to obtain an analytical
representation of the metric functions, which can then be used to study
the stability of such black holes, the motion of particles and fields in
their vicinity~\cite{Zhidenko:2007sj}, or to construct viable
approximations for metrics with incorrect asymptotical behaviour, \eg due
to the presence of magnetic fields
\cite{Konoplya:2007yy,Kokkotas:2010zd}.

As a concluding remark we note that our approach has so far investigated
spherically symmetric spacetimes and hence black holes that are either
nonrotating or slowly rotating. It would be interesting to find a
generalization of our framework for the parametrization of axisymmetric
black holes, for instance, via the application of the Newman-Janis
algorithm. However, while this is technically possible, it is unclear
whether such approach will turn out to be sufficiently robust. We
believe, in fact, that a parametrization of axisymmetric black holes must
combine, together with a rapidly converging expansion, also information
on the parametrized post-Newtonian parameters, on the multipole moments,
on the horizon shape, as well as parameters that define the near-horizon
geometry. This task, which is further complicated by the lack of known
axisymmetric back-hole solutions in alternative theories of gravity (\cf
\cite{Johannsen:2013rqa}), will be the focus of our future work.

\appendix
\section{Johannsen-Psaltis parametrization for the dilaton black hole}
\label{appendix_a}

In order to explore convergence properties of our parametrization
framework, we consider in this appendix the alternative parametrization
of a spherically symmetric black hole in generic metric theories of
gravity which has been recently proposed in Ref.~\cite{Cardoso:2014rha},
\begin{eqnarray}
ds^2&=&-\left[1+h^t(r)\right]\left(1-\frac{2\tilde{M}}{r}\right)dt^2 \nonumber\\
&& +
\left[{1+h^r(r)}\right]\left(1-\frac{2\tilde{M}}{r}\right)^{-1} dr^2 +
r^2 d\Omega^2\,,\label{JP}
\end{eqnarray}
where, instead of the function $h(r)$ in (\ref{JPbh}), two different
functions are introduced:
\begin{subequations}
\label{JPn}
\begin{align}
h^t(r) &\equiv \sum_{n=1}^{\infty}\epsilon_n^t\left(\frac{\tilde{M}}{r}\right)^n
=\epsilon_1^t\frac{\tilde{M}}{r} +\epsilon_2^t\frac{\tilde{M}^2}{r^2} +
\epsilon_3^t\frac{\tilde{M}^3}{r^3} + \ldots\,,\\
h^r(r) &\equiv \sum_{n=1}^{\infty}\epsilon_n^r\left(\frac{\tilde{M}}{r}\right)^n
=\epsilon_1^r\frac{\tilde{M}}{r} +\epsilon_2^r\frac{\tilde{M}^2}{r^2} +
\epsilon_3^r\frac{\tilde{M}^3}{r^3} + \ldots\,.
\end{align}
\end{subequations}

In particular, we will determine the numerical values of the coefficients
$\epsilon_1^t, \epsilon_1^r, \epsilon_2^t, \epsilon_2^r\ldots$ to produce
an approximation of the metric of a dilaton black hole
(\ref{dbh}). Although this can be done in different ways, the
coefficients must obey the constraints that the theory naturally imposes
on them. As a result, the large-distance properties of the functions
$h^t(r)$ and $h^r(r)$ provide a series of constraints that fix the
coefficients once an asymptotic expansion for the metric functions is
made. In this way, we find that [\cf Eqs.~\eqref{JPrel} and
  \eqref{dilaton0}]
\begin{subequations}
\begin{align}
&\tilde{M}=\sqrt{\mu(\mu+b)},\\
&\epsilon_1^t=2-2\sqrt{1+\frac{b}{\mu}},\\
&\epsilon_2^t=\frac{2b}{\mu}+4-4\sqrt{1+\frac{b}{\mu}}\,,\\
&\epsilon_1^r=0\,\ldots\,.
\end{align}
\end{subequations}

\begin{figure}
\resizebox{\linewidth}{!}{ \includegraphics{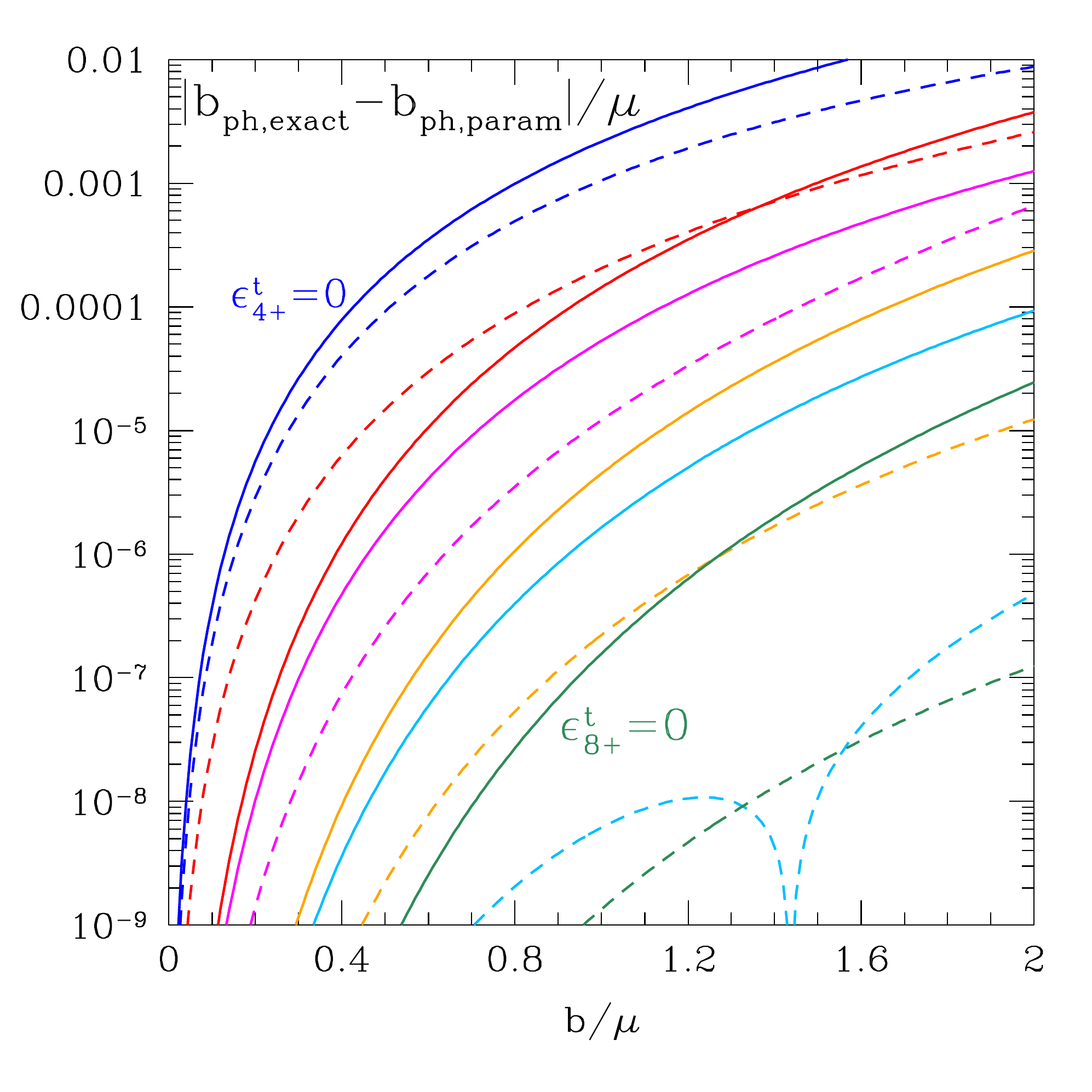}
}
\caption{Difference between the exact values of the
  dilaton black-hole orbit impact parameter for a circular orbit $b_{\rm
    ph}$ and the values obtained using the generalized Johannsen-Psaltis
  metric with the coefficients calculated by comparing the asymptotic
  expansions for the metric functions. The
  results are shown as a function of the dimensionless strength of the
  dilaton parameter $b/\mu$. Different lines refer to different levels of
  approximation, \ie $0=\epsilon_4^t=\epsilon_5^t=\epsilon_6^t=\ldots$
  (blue line), $0=\epsilon_5^t=\epsilon_6^t=\epsilon_7^t=\ldots$ (red
  line), and $0=\epsilon_6^t=\epsilon_7^t=\epsilon_8^t=\ldots$ (magenta
  line), and so on. The dashed lines of the same color correspond to our
  continued-fraction approximation having the same number of parameters;
  hence the first three lines should be compared with those of
  Fig.~\ref{fig:freqdiff} with the same color.}
\label{fig:axion-dilaton-JPn}
\end{figure}

Figure~\ref{fig:axion-dilaton-JPn} shows the relative difference for the
impact parameter of the photon circular relative to a dilaton black hole
as computed using the parametrization~\eqref{JPn}, and shown as function
of the dimensionless strength of the dilaton parameter. Different lines
refer to different levels of approximation, \ie
$0=\epsilon_4^t=\epsilon_5^t=\epsilon_6^t=\ldots$ (blue line),
$0=\epsilon_5^t=\epsilon_6^t=\epsilon_7^t=\ldots$ (red line), and
$0=\epsilon_6^t=\epsilon_7^t=\epsilon_8^t=\ldots$ (magenta line), and so
on. The dashed lines of the same color correspond to our
continued-fraction approximation having the same number of
parameters. More specifically, the first three of these lines should be
compared with the corresponding ones in Fig.~\ref{fig:freqdiff}, in the
following sense: considering, for instance, that the red line in
Fig.~\ref{fig:freqdiff} amounts to specifying four coefficients (\ie
$\epsilon$, $a_0, a_1$, and $a_2$), which is the same number that is
involved when considering the red line in
Fig.~\ref{fig:axion-dilaton-JPn} (\ie $\epsilon_1^t$, $\epsilon_2^t$,
$\epsilon_3^t$, $\epsilon_4^t$). Clearly, the errors in the novel
parametrization are overall smaller for the same number of fixed
coefficients in the expansion. A qualitatively similar figure can be
produced also for the measurement of the ISCO but we do not report it
here for compactness.

We should note that the errors in the parametrization~\eqref{JPn} can be
made smaller if we fix the first two coefficients, \ie $\epsilon_1^t$ and
$\epsilon_2^t$, from the asymptotic expansion at large distance, but we
compute the remaining coefficients from the near-horizon behavior. This
is simply because the impact parameter is a strong-field quantity and
hence its approximation necessarily improves if the coefficients are
constrained near the horizon. On the other hand, the real problematic
feature of the parametrization~\eqref{JPn} is that the coefficients are
roughly equally important near the horizon \cite{Cardoso:2014rha}. As a
result, if one fixes the coefficients by matching the near-horizon
behavior of the metric functions, the expression for the same coefficient
will be different for different orders of approximations, making the
approach not useful for constraining the parameters of the theory.

Finally, as mentioned in the Introduction, a particularly serious
difficulty of the parametrization~\eqref{JPn} is that it does not
reproduce the correct rotating metric even in the regime of slow
rotation. This can be shown rather simply for the slowly rotating
dilaton black hole, for which both the dilaton and the
rotating Johannsen-Psaltis black hole can be obtained with the help of
the Newman-Janis algorithm. More specifically, the generalized
Johannsen-Psaltis black hole in the regime of slow rotation reads
\cite{Cardoso:2014rha}
\begin{eqnarray}
ds^2&=&-\left[1+h^t(r)\right]\left(1-\frac{2\tilde{M}}{r}\right)dt^2 \nonumber\\
&& +
\left[{1+h^r(r)}\right]\left(1-\frac{2\tilde{M}}{r}\right)^{-1} dr^2 +
r^2 d\Omega^2\,\label{JPslow}\\\nonumber
&&-2a\sin^2\theta\left(\sqrt{(1+h^t(r))(1+h^r(r))}\right.\\\nonumber
&&\left.-\left(1-\frac{2\tilde{M}}{r}\right)(1+h^t(r))\right)dt\,d\phi+{\cal O}(a^2).
\end{eqnarray}

By comparing the diagonal elements of (\ref{JPslow}) and (\ref{dabh}) we
conclude that $r^2=\rho^2+2b\rho$, while $h^t(r)$ and $h^r(r)$ must
approximately be such that
\begin{align}
\left(1-\frac{2\tilde{M}}{r}\right)(1+h^t)\simeq&\frac{\rho-2\mu}{\rho+2b}\,,\\
\sqrt{(1+h^t)(1+h^r)}\simeq&\frac{d\rho}{dr}=
\frac{r}{\rho+b}=\frac{\sqrt{\rho^2+2b\rho}}{\rho+b}\,.
\end{align}
As a result, an inconsistency emerges for the off-diagonal element of the
metric (\ref{JPslow}), for which
\begin{align}
\sqrt{(1+h^t)(1+h^r)}-&\left(1-\frac{2\tilde{M}}{r}\right)(1+h^t)%\simeq
\nonumber\\
&\simeq\frac{\sqrt{\rho^2+2b\rho}}{\rho+b}-\frac{\rho-2\mu}{\rho+2b}
\nonumber\\
&\neq\frac{2(\mu+b)}{\rho+2b}\,,
\end{align}
unless $b\ll\rho$. Thus, for any approximation of the functions $h^t(r)$
and $h^r(r)$, the slowly rotating regime of the dilaton black hole is
not reproduced by the metric (\ref{JPslow}). Similar arguments were used
to show that the Newman-Janis algorithm is not able to generate rotating
black-hole solutions in modified gravity theories \cite{Hansen:2013owa}.

\begin{acknowledgments}
It is a pleasure to thank E. Barausse for stimulating discussions and
for providing us with the numerical data from \cite{Barausse:2011pu}.  We
also thank E. Barausse, V. Cardoso, T. Johannsen, and D. Psaltis for
useful comments and suggestions. A.~Z. was supported by the Alexander von
Humboldt Foundation, Germany, and Coordena\c{c}\~ao de Aperfei\c{c}oamento
de Pessoal de N\'ivel Superior (CAPES), Brazil. Partial support comes
from the DFG Grant No. SFB/Transregio~7 and by ``NewCompStar'', COST Action
MP1304.
\end{acknowledgments}


\begin{thebibliography}{99}

\bibitem{Abramowicz:02} M.~A.~Abramowicz, W.~Kluzniak, and J.~P Lasota,
  Astron. Astrophys. {\bf 396} L31-4 (2002)

\bibitem{Doeleman} S.~S.~Doeleman {\it et al.}, Nature, {\bf 455}, 78 (2008).

\bibitem{Falcke:00}
H.~Falcke, F.~Melia, and E.~Agol, Astrophys.\ J.\ Lett., {\bf 528}, L13-16 (2000).

%\cite{Johannsen:2012vz}
\bibitem{Johannsen:2012vz}
  T.~Johannsen, D.~Psaltis, S.~Gillessen, D.~P.~Marrone, F.~Ozel, S.~S.~Doeleman and V.~L.~Fish,
  %``Masses of Nearby Supermassive Black Holes with Very-Long Baseline Interferometry,''
  Astrophys.\ J.\  {\bf 758}, 30 (2012)
  [arXiv:1201.0758 [astro-ph.GA]].
  %%CITATION = ARXIV:1201.0758;%%

%\cite{Bambi:2008jg}
\bibitem{Bambi:2008jg}
  C.~Bambi and K.~Freese,
  %``Apparent shape of super-spinning black holes,''
  Phys.\ Rev.\ D {\bf 79}, 043002 (2009)
  [arXiv:0812.1328 [astro-ph]].
  %%CITATION = ARXIV:0812.1328;%%

\bibitem{JohannsenPsaltis:2010}
%\cite{Johannsen:2010xs}
%\bibitem{Johannsen:2010xs}
  T.~Johannsen and D.~Psaltis,
  %``Testing the No-Hair Theorem with Observations in the Electromagnetic Spectrum: I. Properties of a Quasi-Kerr Spacetime,''
  Astrophys.\ J.\  {\bf 716}, 187 (2010)
  [arXiv:1003.3415 [astro-ph.HE]];
  %%CITATION = ARXIV:1003.3415;%%
%\cite{Johannsen:2010ru}
%\bibitem{Johannsen:2010ru}
%  T.~Johannsen and D.~Psaltis,
  %``Testing the No-Hair Theorem with Observations in the Electromagnetic Spectrum: II. Black-Hole Images,''
  Astrophys.\ J.\  {\bf 718}, 446 (2010)
  [arXiv:1005.1931 [astro-ph.HE]].
  %%CITATION = ARXIV:1005.1931;%%

%\cite{Loeb:2013lfa}
\bibitem{Loeb:2013lfa}
  A.~E.~Broderick, T.~Johannsen, A.~Loeb and D.~Psaltis,
  %``Testing the No-Hair Theorem with Event Horizon Telescope Observations of Sagittarius A*,''
  arXiv:1311.5564 [astro-ph.HE].
  %%CITATION = ARXIV:1311.5564;%%

%\cite{Vigeland:2011ji}
\bibitem{Vigeland:2011ji}
  S.~Vigeland, N.~Yunes and L.~Stein,
  %``Bumpy Black Holes in Alternate Theories of Gravity,''
  Phys.\ Rev.\ D {\bf 83}, 104027 (2011)
  [arXiv:1102.3706 [gr-qc]].
  %%CITATION = ARXIV:1102.3706;%%

%\cite{Will:2005va}
\bibitem{Will:2005va}
  C.~M.~Will,
  %``The Confrontation between general relativity and experiment,''
  Living Rev.\ Rel.\  {\bf 9}, 3 (2006)
  [gr-qc/0510072].
  %%CITATION = GR-QC/0510072;%%

%\cite{Johannsen:2011dh}
\bibitem{Johannsen:2011dh}
  T.~Johannsen and D.~Psaltis,
  %``A Metric for Rapidly Spinning Black Holes Suitable for Strong-Field Tests of the No-Hair Theorem,''
  Phys.\ Rev.\ D {\bf 83}, 124015 (2011)
  [arXiv:1105.3191 [gr-qc]].
  %%CITATION = ARXIV:1105.3191;%%

%\cite{Cardoso:2014rha}
\bibitem{Cardoso:2014rha}
  V.~Cardoso, P.~Pani and J.~Rico,
  %``On generic parametrizations of spinning black-hole geometries,''
  Phys.\ Rev.\ D {\bf 89}, 064007 (2014)
  [arXiv:1401.0528 [gr-qc]].
  %%CITATION = ARXIV:1401.0528;%%

\bibitem{Einstein-Dilaton-Gauss-Bonnet-black-hole}
%\cite{oai:arXiv.org:hep-th/9511071}
%\bibitem{oai:arXiv.org:hep-th/9511071}
  P.~Kanti, N.~E.~Mavromatos, J.~Rizos, K.~Tamvakis and E.~Winstanley,
  %``Dilatonic black holes in higher curvature string gravity,''
  Phys.\ Rev.\ D {\bf 54}, 5049 (1996)
  [hep-th/9511071].
  %%CITATION = HEP-TH/9511071;%%
%\cite{oai:arXiv.org:hep-th/9703192}
%\bibitem{oai:arXiv.org:hep-th/9703192}
  P.~Kanti, N.~E.~Mavromatos, J.~Rizos, K.~Tamvakis and E.~Winstanley,
  %``Dilatonic black holes in higher curvature string gravity. 2: Linear stability,''
  Phys.\ Rev.\ D {\bf 57}, 6255 (1998)
  [hep-th/9703192].
  %%CITATION = HEP-TH/9703192;%%


\bibitem{Aether}
  R.~A.~Konoplya and A.~Zhidenko,
  %``Perturbations and quasi-normal modes of black holes in Einstein-Aether theory,''
  Phys.\ Lett.\ B {\bf 644}, 186 (2007)
  [gr-qc/0605082];
  %%CITATION = GR-QC/0605082;%%
  %``Gravitational spectrum of black holes in the Einstein-Aether theory,''
  Phys.\ Lett.\ B {\bf 648}, 236 (2007)
  [hep-th/0611226].
  %%CITATION = HEP-TH/0611226;%%

\bibitem{Garcia:1995}
  A.~Garcia, D.~Galtsov and O.~Kechkin,
  %``Class of stationary axisymmetric solutions of the Einstein-Maxwell dilaton - axion field equations,''
  Phys.\ Rev.\ Lett.\  {\bf 74}, 1276 (1995).
  %%CITATION = PRLTA,74,1276;%%

%\cite{Wei:2013kza}
\bibitem{Wei:2013kza}
  S.~-W.~Wei and Y.~-X.~Liu,
  %``Observing the shadow of Einstein-Maxwell-Dilaton-Axion black hole,''
  JCAP {\bf 1311}, 063 (2013)
  [arXiv:1311.4251 [gr-qc]].
  %%CITATION = ARXIV:1311.4251;%%

\bibitem{JacobsonAE}
%\cite{Jacobson:2000xp}
%\bibitem{Jacobson:2000xp}
  T.~Jacobson and D.~Mattingly,
  %``Gravity with a dynamical preferred frame,''
  Phys.\ Rev.\ D {\bf 64}, 024028 (2001)
  [gr-qc/0007031];
  %%CITATION = GR-QC/0007031;%%
%\cite{Jacobson:2008aj}
%\bibitem{Jacobson:2008aj}
  T.~Jacobson,
  %``Einstein-aether gravity: A Status report,''
  PoS QG {\bf -PH}, 020 (2007)
  [arXiv:0801.1547 [gr-qc]].
  %%CITATION = ARXIV:0801.1547;%%

%\cite{Barausse:2011pu}
\bibitem{Barausse:2011pu}
  E.~Barausse, T.~Jacobson and T.~P.~Sotiriou,
  %``Black holes in Einstein-aether and Horava-Lifshitz gravity,''
  Phys.\ Rev.\ D {\bf 83}, 124043 (2011)
  [arXiv:1104.2889 [gr-qc]].
  %%CITATION = ARXIV:1104.2889;%%

\bibitem{Yagietal_13}
%\cite{Yagi:2013qpa}
%\bibitem{Yagi:2013qpa}
  K.~Yagi, D.~Blas, N.~Yunes and E.~Barausse,
  %``Strong Binary Pulsar Constraints on Lorentz Violation in Gravity,''
  Phys.\ Rev.\ Lett.\  {\bf 112}, 161101 (2014)
  [arXiv:1307.6219 [gr-qc]];
  %%CITATION = ARXIV:1307.6219;%%
%\cite{Yagi:2013ava}
%\bibitem{Yagi:2013ava}
%  K.~Yagi, D.~Blas, E.~Barausse and N.~Yunes,
  %``Constraints on Einstein-\AE ther theory and Horava gravity from binary pulsar observations,''
  Phys.\ Rev.\ D {\bf 89}, 084067 (2014)
  [arXiv:1311.7144 [gr-qc]].
  %%CITATION = ARXIV:1311.7144;%%

%\cite{Konoplya:2011qq}
\bibitem{Konoplya:2011qq}
  R.~A.~Konoplya and A.~Zhidenko,
  %``Quasinormal modes of black holes: From astrophysics to string theory,''
  Rev.\ Mod.\ Phys.\  {\bf 83}, 793 (2011)
  [arXiv:1102.4014 [gr-qc]].
  %%CITATION = ARXIV:1102.4014;%%

%\cite{Barausse:2014tra}
\bibitem{Barausse:2014tra}
  E.~Barausse, V.~Cardoso and P.~Pani,
  %``Can environmental effects spoil precision gravitational-wave astrophysics?,''
  Phys.\ Rev.\ D {\bf 89}, 104059 (2014)
  [arXiv:1404.7149 [gr-qc]].
  %%CITATION = ARXIV:1404.7149;%%

%\cite{Gundlach:1993tp}
\bibitem{Gundlach:1993tp}
  C.~Gundlach, R.~H.~Price, and J.~Pullin,
  %``Late time behavior of stellar collapse and explosions: 1. Linearized
  %perturbations,''
  Phys.\ Rev.\  D {\bf 49}, 883 (1994)
 [arXiv:gr-qc/9307009].

\bibitem{Ryan:1995-1997}
%\cite{Ryan:1995wh}
%\bibitem{Ryan:1995wh}
  F.~D.~Ryan,
  %``Gravitational waves from the inspiral of a compact object into a massive, axisymmetric body with arbitrary multipole moments,''
  Phys.\ Rev.\ D {\bf 52}, 5707 (1995);
  %%CITATION = PHRVA,D52,5707;%%
%\cite{Ryan:1996nk}
%\bibitem{Ryan:1996nk}
%  F.~D.~Ryan,
  %``Spinning boson stars with large selfinteraction,''
  Phys.\ Rev.\ D {\bf 55}, 6081 (1997);
  %%CITATION = PHRVA,D55,6081;%%
%\cite{Ryan:1997kh}
%\bibitem{Ryan:1997kh}
%  F.~D.~Ryan,
  %``Scalar waves produced by a scalar charge orbiting a massive body with arbitrary multipole moments,''
  Phys.\ Rev.\ D {\bf 56}, 7732 (1997).
  %%CITATION = PHRVA,D56,7732;%%
%\cite{Collins:2004ex}
\bibitem{Collins:2004ex}
  N.~A.~Collins and S.~A.~Hughes,
  %``Towards a formalism for mapping the space-times of massive compact objects: Bumpy black holes and their orbits,''
  Phys.\ Rev.\ D {\bf 69}, 124022 (2004)
  [gr-qc/0402063].
  %%CITATION = GR-QC/0402063;%%
%\cite{Glampedakis:2005cf}
\bibitem{Glampedakis:2005cf}
  K.~Glampedakis and S.~Babak,
  %``Mapping spacetimes with LISA: Inspiral of a test-body in a `quasi-Kerr' field,''
  Class.\ Quant.\ Grav.\  {\bf 23}, 4167 (2006)
  [gr-qc/0510057].
  %%CITATION = GR-QC/0510057;%%

%\cite{Barausse:2012qh}
\bibitem{Barausse:2012qh}
  E.~Barausse and T.~P.~Sotiriou,
  %``Slowly rotating black holes in Horava-Lifshitz gravity,''
  Phys.\ Rev.\ D {\bf 87}, 087504 (2013)
  [arXiv:1212.1334].
  %%CITATION = ARXIV:1212.1334;%%
%\cite{Wang:2012nv}
%\bibitem{Wang:2012nv}
  A.~Wang,
  %``Stationary and slowly rotating spacetimes in Ho\v{r}ava-Lifshitz gravity,''
  Phys.\ Rev.\ Lett.\  {\bf 110}, 091101 (2013)
  [arXiv:1212.1876].
  %%CITATION = ARXIV:1212.1876;%%


%\cite{Barausse:2013nwa}
\bibitem{Barausse:2013nwa}
  E.~Barausse and T.~P.~Sotiriou,
  %``Black holes in Lorentz-violating gravity theories,''
  Class.\ Quant.\ Grav.\  {\bf 30} (2013) 244010
  [arXiv:1307.3359 [gr-qc]].
  %%CITATION = ARXIV:1307.3359;%%

%\cite{Konno:2009kg}
\bibitem{Konno:2009kg}
  K.~Konno, T.~Matsuyama and S.~Tanda,
  %``Rotating black hole in extended Chern-Simons modified gravity,''
  Prog.\ Theor.\ Phys.\  {\bf 122}, 561 (2009)
  [arXiv:0902.4767 [gr-qc]].
  %%CITATION = ARXIV:0902.4767;%%
%\cite{Yunes:2009hc}
%\bibitem{Yunes:2009hc}
  N.~Yunes and F.~Pretorius,
  %``Dynamical Chern-Simons Modified Gravity. I. Spinning Black Holes in the Slow-Rotation Approximation,''
  Phys.\ Rev.\ D {\bf 79}, 084043 (2009)
  [arXiv:0902.4669 [gr-qc]];
  %%CITATION = ARXIV:0902.4669;%%
%\cite{Yagi:2012ya}
%\bibitem{Yagi:2012ya}
  K.~Yagi, N.~Yunes and T.~Tanaka,
  %``Slowly Rotating Black Holes in Dynamical Chern-Simons Gravity: Deformation Quadratic in the Spin,''
  Phys.\ Rev.\ D {\bf 86}, 044037 (2012)
  [arXiv:1206.6130 [gr-qc]].
  %%CITATION = ARXIV:1206.6130;%%

%\cite{Pani:2009wy}
\bibitem{Pani:2009wy}
  P.~Pani and V.~Cardoso,
  %``Are black holes in alternative theories serious astrophysical candidates? The Case for Einstein-Dilaton-Gauss-Bonnet black holes,''
  Phys.\ Rev.\ D {\bf 79}, 084031 (2009)
  [arXiv:0902.1569 [gr-qc]].
  %%CITATION = ARXIV:0902.1569;%%
%\cite{Pani:2011gy}
%\bibitem{Pani:2011gy}
  P.~Pani, C.~F.~B.~Macedo, L.~C.~B.~Crispino and V.~Cardoso,
  %``Slowly rotating black holes in alternative theories of gravity,''
  Phys.\ Rev.\ D {\bf 84}, 087501 (2011)
  [arXiv:1109.3996 [gr-qc]].
  %%CITATION = ARXIV:1109.3996;%%
%\cite{Ayzenberg:2014aka}
%\bibitem{Ayzenberg:2014aka}
  D.~Ayzenberg and N.~Yunes,
  %``Slowly-Rotating Black Holes in Einstein-Dilaton-Gauss-Bonnet Gravity: Quadratic Order in Spin Solutions,''
  Phys.\ Rev.\ D {\bf 90}, 044066 (2014)
  [arXiv:1405.2133 [gr-qc]].
  %%CITATION = ARXIV:1405.2133;%%


\bibitem{NJ65}
  E.~T. Newman and A.~I. Janis, J.\ Math.\ Phys.\ (N.Y.)  {\bf 6}, 915 (1965);
  S.~P. Drake and P. Szekeres, Gen.\ Relativ.\ Gravit. {\bf 32}, 445 (2000).

%\cite{Yazadjiev:1999ce}
\bibitem{Yazadjiev:1999ce}
  S.~Yazadjiev,
  %``Newman-Janis method and rotating dilaton axion black hole,''
  Gen.\ Rel.\ Grav.\  {\bf 32}, 2345 (2000)
  [gr-qc/9907092].
  %%CITATION = GR-QC/9907092;%%

\bibitem{Zhidenko:2007sj}
  A.~Zhidenko,
  %``Quasi-normal modes for black hole solutions unknown in analytical form,''
  arXiv:0705.2254 [gr-qc].
  %%CITATION = ARXIV:0705.2254;%%
%\cite{Konoplya:2007yy}
\bibitem{Konoplya:2007yy}
  R.~A.~Konoplya and R.~D.~B.~Fontana,
  %``Quasinormal modes of black holes immersed in a strong magnetic field,''
  Phys.\ Lett.\ B {\bf 659}, 375 (2008)
  [arXiv:0707.1156 [hep-th]].
  %%CITATION = ARXIV:0707.1156;%%
%\cite{Kokkotas:2010zd}
\bibitem{Kokkotas:2010zd}
  K.~D.~Kokkotas, R.~A.~Konoplya and A.~Zhidenko,
  %``Quasinormal modes, scattering and Hawking radiation of Kerr-Newman black holes in a magnetic field,''
  Phys.\ Rev.\ D {\bf 83}, 024031 (2011)
  [arXiv:1011.1843 [gr-qc]].
  %%CITATION = ARXIV:1011.1843;%%

%\cite{Johannsen:2013rqa}
\bibitem{Johannsen:2013rqa}
  T.~Johannsen,
  %``Systematic Study of Event Horizons and Pathologies of Parametrically Deformed Kerr Spacetimes,''
  Phys.\ Rev.\ D {\bf 87}, no. 12, 124017 (2013)
  [arXiv:1304.7786 [gr-qc]].
  %%CITATION = ARXIV:1304.7786;%%

%\cite{Hansen:2013owa}
\bibitem{Hansen:2013owa}
  D.~Hansen and N.~Yunes,
  %``Applicability of the Newman-Janis Algorithm to Black Hole Solutions of Modified Gravity Theories,''
  Phys.\ Rev.\ D {\bf 88}, no. 10, 104020 (2013)
  [arXiv:1308.6631 [gr-qc]].
  %%CITATION = ARXIV:1308.6631;%%
\end{thebibliography}
\end{document}